\documentclass[pra,showpacs,onecolumn,10pt]{revtex4-1}
\usepackage{indentfirst}
\usepackage{graphicx}
\usepackage{amsmath}
\usepackage{amsfonts}
\usepackage{color}
\usepackage[sc]{mathpazo}
\begin{document}
\title{Correlation and nonlocality measures as indicators of quantum phase transitions in several critical systems}
\author{Ferdi Altintas}\email{ferdialtintas@ibu.edu.tr}
\author{Resul Eryigit}\email{resul@ibu.edu.tr}
\affiliation{Department of Physics, Abant Izzet Baysal University, Bolu, 14280, Turkey.}
\begin{abstract}
We have investigated the quantum phase transitions in the ground states of several critical systems, including transverse field Ising and XY models as well as XY with multiple spin interactions, XXZ and the collective system Lipkin-Meshkov-Glick models, by using different quantumness measures, such as entanglement of formation, quantum discord, as well as its classical counterpart, measurement-induced disturbance and the Clauser-Horne-Shimony-Holt-Bell function. Measurement-induced disturbance is found to detect the first and second order phase transitions present in these critical systems, while, surprisingly, it is found to fail to signal the infinite-order phase transition present in the XXZ model. Remarkably, the Clauser-Horne-Shimony-Holt-Bell function is found to detect all the phase transitions, even when quantum and classical correlations are zero for the relevant ground state.
\end{abstract}


\maketitle
\section{Introduction}
Classical phase transitions occur at a finite critical temperature at which the physical system reaches a state characterized by macroscopic order~\cite{cpt}. Unlike classical phase transitions, quantum phase transitions~(QPT), occur at absolute zero temperature as a tuning parameter~(a coupling constant or an external parameter) of the system's Hamiltonian is changed and are solely due to the quantum fluctuations~\cite{sachdev}. The superfluid-Mott insulator transition in a Bose-Einstein condensate~\cite{greiner}, superconductor-insulator transitions~\cite{vgvd} and the phase transition in spin models at $T=0$~\cite{afov} are the remarkable examples of the QPTs. Although the quantum and classical phase transitions are both characterized by the specific behavior of correlation length in the system, the quantum correlations, mostly, have no classical counterparts. Different measures, such as entanglement, quantum discord~(QD) and violation of Bell inequalities, have been introduced to account for the "quantumness" of a given state and it was shown that such measures can be used as a fundamental resource in the protocols of quantum informational and computational tasks~\cite{qsuproto}. Several attempts have been made to find a universal quantum correlation measure that can unambiguously signal all the QPTs and to search for a relation between QPTs and quantum correlation measures~\cite{lwmsdl,eryigit,ccrmss,tomn,wsls,sarandy,lslzw,mgcss,wtrr,wrr,dillenschneider,ylhl,crms,plaqpt,pznp,wcjl,jbmc,ljto,dwcgyo}. Indeed, different measures of quantum correlations, such as entanglement~\cite{lwmsdl,eryigit,ccrmss,tomn,wsls}, quantum discord~\cite{sarandy,lslzw,mgcss,wtrr,wrr,dillenschneider,ylhl,crms,plaqpt}, fidelity~\cite{pznp,wcjl} and Bell inequalities~\cite{jbmc,ljto,dwcgyo} have been shown to provide the signatures of QPTs. Although not always, the extremal points~(minimums or maximums) as well as the behavior of the derivatives of the correlation measures can signal the existence of  a critical point~(CP). In general, at the first-order QPT point~\cite{orderqpt}, QD  and entanglement~(measured by concurrence and entanglement of formation~(EoF)) are found to be discontinuous at the critical point~\cite{sarandy,wtrr,wrr}, while at the second-order QPT, the divergence or discontinuity behavior of the  first and second derivatives of QD and entanglement could spotlight the critical points~\cite{sarandy,lslzw,mgcss}. On the other hand, QD and entanglement are found to show local extreme points at the CPs of infinite-order QPTs~\cite{wtrr,wrr,dillenschneider}.

On the other hand, entanglement may fail to measure the quantum correlations of a state when it reaches to a certain level of mixture, especially for distant spin pairs in a spin system and at finite temperatures. In addition, QPTs might involve transitions from unentangled to unentangled ground states. On the other hand, quantum discord, mostly, does not vanish for mixed separable states, actually almost all quantum states have non-zero quantum discord~\cite{ferraro}. This feature makes QD more reliable and useful measure compared to entanglement to signal the QPTs. Indeed, QD is able to detect the critical points of QPTs, especially between far neighbors and also at finite temperatures where entanglement is fully absent near the critical points~\cite{mgcss,wtrr,wrr,ylhl}. Moreover, QD can provide more conventional information about the critical points of QPTs than that of the conventional measures~(thermodynamical quantities, such as entropy, specific heat, magnetization, etc.); as shown recently QD, in contrast to entanglement and some of the thermodynamics quantities, can spotlight the QPTs, even at finite temperatures~\cite{wtrr,wrr}.  It should be pointed out here that  Bell nonlocality as measured by the Clauser-Horne-Shimony-Holt-type~(CHSH) inequality or its derivative has been used to detect the quantum phase transitions only in a few model systems, including Heisenberg XXZ~\cite{ljto} and Kitaev-Castelnovo-Chamon models~\cite{dwcgyo}.  In these studies, it was shown that CHSH-Bell inequality, although not violated for some cases, exhibits the existence of QPTs present in the models, even for separable states. Moreover, in Ref.~\cite{jbmc}, the authors investigated  only the behavior of the CHSH-Bell function near the Ising-like critical point in the infinite XY model and found a non-ergodic behavior of the nonlocality measure near the phase transition point.  However, making a general conclusion about the ability of the CHSH inequality to detect the CPs of QPTs is hard.  On the other hand, although a number of studies indicate a strong relation between the QPT and a discontinuity in QD and entanglement or in their derivatives, so far there is no single correlation measure that can indicate all the QPTs or shows universal behavior at the CPs of QPTs.

The main motivation of the present study is to search for new quantum information theory based indicator of quantum phase transitions, such as measurement-induced disturbance~(MID) and the CHSH-Bell function, and compare them with the well known CP-detectors, quantum discord, classical correlations~(CC) and entanglement. Towards that goal, we will investigate different correlation measures, such as EoF, QD and its classical counterpart as well as the CHSH-Bell function and the quantumness measure, the so-called measurement-induced disturbance~(the explicit definitions of each correlation measure will be provided later in the text) in the ground states of several critical systems, including transverse field Ising and XY models as well as the XY model with multiple spin interactions, the XXZ model and the collective system Lipkin-Meshkov-Glick~(LMG) model. The focus of the present study will be two fold. On one hand, we will compare and contrast the behavior of different quantumness measures as a function of the tuning parameter of the system Hamiltonians, especially for the regions close to the critical points of the QPTs of the system. On the other hand, we will study different critical systems that exhibit first and infinite order QPTs as well as several continuous phase transitions belonging to different universality classes to see whether or not the corresponding correlation measures~(particularly MID and the CHSH-Bell function) can signal the abrupt change of the ground state. We will assume both finite and infinite chains as well as the finite size effects of the derivatives of the correlations at the second order QPT points. An analysis of QD and EoF for these critical systems has been recently reported in Refs.~\cite{sarandy,mgcss,ylhl}. Here we will both extend and review the results in Refs.~\cite{sarandy,mgcss,ylhl}, and also introduce MID and CHSH-Bell function as new theoretical CP-detectors for these systems. In fact, by considering a number of models that exhibit several  QPTs obeying different universality classes, with different correlation measures, we will shed some light on the search for "universal" correlation measure that can account for all the QPTs. We have shown that although none of the considered correlation measures show a universal behavior for signaling the CPs of QPTs, remarkably, the CHSH-Bell function is found to detect all CPs of QPTs present in the considered models, even when the relevant ground state has neither quantum nor classical correlations.

The paper is organized as follows. In Sec.~\ref{sec1}, we review the correlation measures for bipartite quantum systems and then give the explicit analytic expressions for general X-states~(a class of states that includes $Z_2$ symmetric states). In Sec.~\ref{sec2}, we analyze the quantum criticality in Ising, XY and XY with multiple spin interaction models. In Secs.~\ref{sec3} and~\ref{sec4}, we focus on the ground state correlations and criticality in XXZ and LMG models, respectively. A brief summary of important results are given in Sec.~\ref{sec5}.
\section{Correlations in Bipartite Quantum Systems}\label{sec1}
As is well known the correlation measures, entanglement, QD, MID, and the CHSH-Bell inequality, signify  different aspects of quantum correlations in a given state. In this section, we will briefly describe what they signify for the "quantumness" of correlations  and then we will give their analytic expressions in terms of density matrix elements for a general two-qubit $X$-structured density matrix.

For the model systems that we will consider in the next sections, the Hamiltonians exhibit $Z_2$ symmetry; the Hamiltonians are invariant under $\pi$ rotation around a given spin axis~\cite{tomn,sarandy}. This symmetry leads to the two-qubit reduced density matrix at sites $i$ and $j$ in the standard basis $\left|1\right\rangle= \left|\uparrow\uparrow\right\rangle, \left|2\right\rangle= \left|\uparrow\downarrow\right\rangle, \left|3\right\rangle= \left|\downarrow\uparrow\right\rangle$ and $\left|4\right\rangle= \left|\downarrow\downarrow\right\rangle$ (where $\left|\uparrow\right\rangle$ and $\left|\downarrow\right\rangle$ are the eigenstates of the Pauli spin 1/2 $z$-operator) has an $X$-structure~\cite{tomn,sarandy}:
\begin{eqnarray}\label{xstate}
\rho_{ij}=\left (\begin{array}{cccc} u_{11}  & 0 & 0  & u_{14} \\ 0  & u_{22} & u_{23}  & 0 \\ 0  & u_{32} & u_{33}  & 0 \\ u_{41}  & 0 & 0  & u_{44} \end{array} \right) \ .
\end{eqnarray}
For such states, the correlation measures can be calculated analytically.

We will use entanglement of formation as a measure of entanglement which is defined as~\cite{wye}:
\begin{eqnarray}\label{eof}
EoF(\rho_{ij})=H\left[\frac{1+\sqrt{1-\mathbb{C}(\rho_{ij})^2}}{2}\right],
\end{eqnarray}
where $H[x]=-x\log_2x-(1-x)\log_2(1-x)$ is the binary entropy function and $\mathbb{C}(\rho_{ij})$ is the concurrence, an entanglement monotone~\cite{wye}. EoF quantifies a class of quantum correlations that cannot be created by local operations and classical communication only~\cite{wye,werner}. It defines whether or not the given bipartite state can be written in terms of the tensor product of states of the subsystems, $i$ and $j$. It is equal to zero for separable states and one for maximally entangled~(Bell) states. For the density matrix~(\ref{xstate}), the concurrence can be obtained easily
\begin{eqnarray}
\mathbb{C}(\rho_{ij})=2\max\{0,|u_{14}|-\sqrt{u_{22}u_{33}}, |u_{23}|-\sqrt{u_{11}u_{44}}\}.\nonumber
\end{eqnarray}

On the other hand, there exist more-general-than-entanglement-type quantum correlations that may exist even for mixed separable states. Recently, Ollivier and Zurek introduced the notion, called quantum discord, as a measure of such more-general-than-entanglement-type quantum correlations for bipartite states~\cite{howz}. Indeed, QD can be non-zero for some mixed separable states. It is defined as the difference between the quantum versions of two classically equivalent definitions of mutual information and can be given as
\begin{eqnarray}\label{qd}
QD(\rho_{ij})=I(\rho_{ij})-CC(\rho_{ij}),
\end{eqnarray}
where $I(\rho_{ij})=S(\rho_i)+S(\rho_j)-S(\rho_{ij})$ is the quantum mutual information and measures the total correlations. Here $\rho_{i,j}=Tr_{j,i}\rho_{ij}$ are the reduced density matrices and $S(\rho)=-Tr(\rho\log_2\rho)$ is the von-Neumann entropy. $CC(\rho_{ij})$ in Eq.~(\ref{qd}) is the measure of classical correlations between two subsystems~\cite{lhvv}. It is defined as the maximum information about one system that can be obtained by performing a set of projective measurements on the other subsystem. Its calculation is based on complex maximization procedure and obtaining an analytic expression of  $CC(\rho_{ij})$  for general states is not an easy task. However, for $X$-states the analytic expression of  $CC(\rho_{ij})$  is available if we restrict ourselves to the projective positive operator valued measurements performed locally on the subsystem $j$~\cite{wlnl}. According to the results in Ref.~\cite{wlnl}, QD and CC are given as:
\begin{eqnarray}\label{qdcc}
QD(\rho_{ij})=\min\{Q_1,Q_2\},\quad CC(\rho_{ij})=\max\{CC_1,CC_2\},
\end{eqnarray}
where $CC_j=H[u_{11}+u_{22}]-D_j$ and $Q_j=H[u_{11}+u_{33}]+\sum_{k=1}^4\lambda^k\log_2\lambda^k+D_j$ with $\lambda^k$ being the eigenvalues of $\rho_{ij}$ and $H[x]$ is the binary entropy defined above. Here $D_1=H[\tau]$ where $\tau=\left(1+\sqrt{[1-2(u_{33}+u_{44})]^2+4(|u_{14}|+|u_{23}|)^2}\right)/2$ and $D_2=-\sum_{k=1}^4 u_{kk}\log_2u_{kk}-H[u_{11}+u_{33}]$.
In general, QD and CC are not symmetric quantities, i.e., their value can depend on the party which projective measurements are performed on. Indeed, for general $X$ states, the order of measurements performed are important since $S(\rho_i)\neq S(\rho_j)$. However for the considered model systems in the present study~(except LMG model), $u_{22}=u_{33}$ which ensures  $S(\rho_i)=S(\rho_j)$; irrespective of whether the measurement is performed locally on the subsystem $i$ or $j$.

Along similar lines, the so-called measurement-induced disturbance was proposed very recently by Luo in order to quantify the more-general-than-entanglement-type quantum correlations~\cite{luo}. If the mutual information, $I(\rho)$, is accepted to quantify all kind of correlations for a bipartite state, then MID is just the mutual information difference between a given state, $\rho$, and the closest classical state to $\rho$.  The idea in the definition of MID can be summarized as follows. For a given state $\rho$, any complete local projective measurements $\{\Pi_i^a\otimes\Pi_j^b\}$ on the subsystems $a$ and $b$  will induce a classical state $\Pi(\rho)$. If the measurements $\{\Pi_i\}$ are chosen as the spectral resolutions of the reduced states; $\rho_a=\sum_ip_i^a\Pi_i^a$ and $\rho_b=\sum_jp_i^b\Pi_j^b$~(where $p_i^{a,b}$ are the probability amplitudes), the state $\Pi(\rho)=\sum_{i,j}(\Pi_i^a\otimes\Pi_j^b)\rho(\Pi_i^a\otimes\Pi_j^b)$ will be the closest classical state to $\rho$. Indeed, the measurements induced by the spectral resolution of the reduced states will leave the marginal states invariant; in the sense of the least disturbing measurements. Then MID is defined as:
\begin{eqnarray}\label{mýd1}
MID=I(\rho)-I(\Pi(\rho)).
\end{eqnarray}
MID is easily computable and can be applied to bipartite higher dimensions. For the density matrix in Eq.~(\ref{xstate}), the closest classical state would be $\Pi(\rho_{ij})=diag\{u_{11},u_{22},u_{33},u_{44}\}$. Since the reduced states are not affected by these measurements and the closest classical state is diagonal, MID reduces to the entropy distance, $S(x||y)=-tr(x\log_2y)-S(x)$, between the given state and the closest classical state to $\rho_{ij}$:
\begin{eqnarray}\label{mýd2}
MID(\rho_{ij})&=&S(\rho_{ij}||\Pi(\rho_{ij}))\nonumber\\
&=&\sum_{k=1}^4\lambda^k\log_2\lambda^k-\sum_{k=1}^4u_{kk}\log_2u_{kk},
\end{eqnarray}
where $\lambda^k$ are the eigenvalues of $\rho_{ij}$.

The last correlation measure that we will use in the present study is the CHSH-Bell inequality~\cite{chsh}. The violation of CHSH inequality describes the part of entanglement~(often called nonlocal correlations) that cannot be reproduced by a local hidden variable model, i.e., by classical systems. Although entangled pure states violate the CHSH inequality, mixed entangled states may not imply that their correlations cannot be classically reproduced with certainty. By using the Horodecki criteria~\cite{hhh}, the maximum of the CHSH Bell function for bipartite X-states can be obtained easily~\cite{djmbfc}:
\begin{eqnarray}\label{bell}
CHSH(\rho_{ij})=\max\{B_1,B_2\},
\end{eqnarray}
where $B_1=2\sqrt{k_1^2+4k_2^2}$ and $B_2=4\sqrt{k_2^2+k_3^2}$, $k_1=(u_{11}+u_{44}-u_{22}-u_{33})$, $k_2=(|u_{14}|+|u_{23}|)$ and $k_3=(|u_{14}|-|u_{23}|)$. When the value of CHSH function exceeds 2 for a given state , it is violated, that is no classical local model can reproduce all correlations of these states.

The correlation measures (EoF, MID and  QD) are equal to each other for bipartite pure states, but they disagree on the value of correlations for mixed states. MID and QD may give non-zero values for some mixed separable states. On the other hand, MID can be nonzero, even maximal for purely classical states~(where $QD=0$). Therefore, MID is sometimes regarded as an unfaithful and nonrefined measure of nonclassical correlations in bipartite states~\cite{gpa,wpm,sclmmp}. The CHSH-Bell function attains its maximum value for maximally entangled states, but in general it is violated for states having high entanglement.

\section{Model Systems, Quantum Phase Transitions and Ground State Correlations}
\subsection{XYT model and its variants}\label{sec2}
The anisotropic spin $1/2$ XY chain with three spin interaction under external transverse field~(XYT model) can be defined by the following Hamiltonian~\cite{ylhl,lzxt}:
\begin{eqnarray}\label{xythamiltonian}
H_{XYT}&=&-\sum_{j=1}^N\left\{\left(\frac{1+\gamma}{2}\right)\sigma_j^x\sigma_{j+1}^x+\left(\frac{1-\gamma}{2}\right)\sigma_j^y\sigma_{j+1}^y+\lambda\sigma_j^z\right\}\nonumber\\
&-&\sum_{j=1}^N\alpha\left(\sigma_{j-1}^x\sigma_{j}^z\sigma_{j+1}^x+\sigma_{j-1}^y\sigma_{j}^z\sigma_{j+1}^y\right),
\end{eqnarray}
where $\sigma_j^{x,y,z}$ are the spin $1/2$ Pauli matrices on site $j$, $N$ is the total number of spins in the chain, $\gamma$ is the degree of anisotropy and $\lambda$ and $\alpha$ are the strengths of the external field and the three spin interaction, respectively.

The XYT model reduces to the transverse field Ising model for $\alpha=0$ and $\gamma=1$ and to the anisotropic XY model for $\alpha=0$. It is exactly solvable and can be diagonalized by mapping it into spinless free fermion model with single orbitals. This can be implemented through the Jordan-Wigner map followed by a Bogoliubov transformation~(see, Refs.~\cite{ylhl,lzxt} for the diagonalization of an XYT Hamiltonian).

The symmetries of the Hamiltonian of the system simplify the structure of the two point reduced density matrix. Since the Hamiltonian of the XYT model is real, then $\rho_{ij}=\rho_{ij}^*$. Because of the translational invariance of $H_{XYT}$, the two-point density matrix  of the system is position independent, i.e., $\rho_{ij}=\rho_{ii+r}$ for integer $r$. Furthermore, the density matrix commutes with $\sigma_i^z\sigma_j^z$ because of the global rotation symmetry~($Z_2$-symmetry) of $H_{XYT}$. As a result of these symmetries, the two qubit reduced density matrix of the system has an X-structure with the following matrix elements~\cite{ylhl}: 
\begin{eqnarray}\label{xytmatelements}
u_{11}&=&\frac{1}{4}\left(1+2\left\langle\sigma^z\right\rangle+\left\langle\sigma_i^z\sigma_{i+r}^z\right\rangle\right),\nonumber\\
u_{22}&=&u_{33}=\frac{1}{4}\left(1-\left\langle\sigma_i^z\sigma_{i+r}^z\right\rangle\right),\nonumber\\
u_{23}&=&u_{32}=\frac{1}{4}\left(\left\langle\sigma_i^x\sigma_{i+r}^x\right\rangle+\left\langle\sigma_i^y\sigma_{i+r}^y\right\rangle\right),\nonumber\\
u_{14}&=&u_{41}=\frac{1}{4}\left(\left\langle\sigma_i^x\sigma_{i+r}^x\right\rangle-\left\langle\sigma_i^y\sigma_{i+r}^y\right\rangle\right),\nonumber\\
u_{44}&=&\frac{1}{4}\left(1-2\left\langle\sigma^z\right\rangle+\left\langle\sigma_i^z\sigma_{i+r}^z\right\rangle\right).
\end{eqnarray}

In the following, we will restrict ourselves to the ground state correlations~(where the temperature $T=0$) between the nearest-neighbor sites~($r=1$). The transverse magnetization $\left\langle\sigma^z\right\rangle$ and the two-point correlation functions between nearest-neighbor sites at $T=0$ for the XYT model can be calculated as~\cite{ebbm1,ebbm2}:
\begin{eqnarray}\label{tmag}
\left\langle\sigma^z\right\rangle&=&\frac{1}{N}\sum_k\frac{\epsilon_k}{\omega_k},\nonumber\\
\left\langle\sigma_i^x\sigma_{i+1}^x\right\rangle&=&G_{-1},\quad \left\langle\sigma_i^y\sigma_{i+1}^y\right\rangle=G_1,\nonumber\\
\left\langle\sigma_i^z\sigma_{i+1}^z\right\rangle&=&\left\langle\sigma^z\right\rangle^2-G_1G_{-1},
\end{eqnarray}
where $G_k=-\sum_k\left(\cos(x_kr)\epsilon_k+\gamma\sin(x_kr)\sin(x_k)\right)/N\omega_k$ and $\omega_k=\sqrt{\epsilon_k^2+(\gamma\sin(x_k))^2}$, $\epsilon_k=\lambda-\cos(x_k)-2\alpha\cos(2x_k)$ is the energy spectrum with $x_k=\frac{2\pi k}{N}$, $k=-M,\ldots ,M$ and $M=(N-1)/2$ for odd $N$.

Now, we discuss the critical behavior of correlation measures for the transverse field Ising, XY and XYT models.  

\subsubsection{Transverse field Ising model}
As a first investigation, we set $\alpha=0$ and $\gamma=1$ in the XYT Hamiltonian which reduces to the well known transverse field Ising Hamiltonian~\cite{dillenschneider}:
\begin{eqnarray}\label{ising}
H_{Ising}=-\sum_{j=1}^N\{\sigma_j^x\sigma_{j+1}^x+\lambda\sigma_j^z\}.
\end{eqnarray}
The system will undergo a second-order quantum phase transition at the critical value $\lambda_c=1$ where the global phase flip symmetry breaks and the correlation length diverges~\cite{afov}. Our aim in this section is two-fold; first we want to analyze the behavior of correlation measures~(EoF, QD, CC, MID and CHSH-Bell function) near the critical point. Second, we want to check the ability of the correlation measures to detect the QPT and investigate the finite-size scaling behavior of the derivatives of the correlations at the critical point, $\lambda_c$.
\begin{figure}[!ht]\centering
\includegraphics[width=7.9cm]{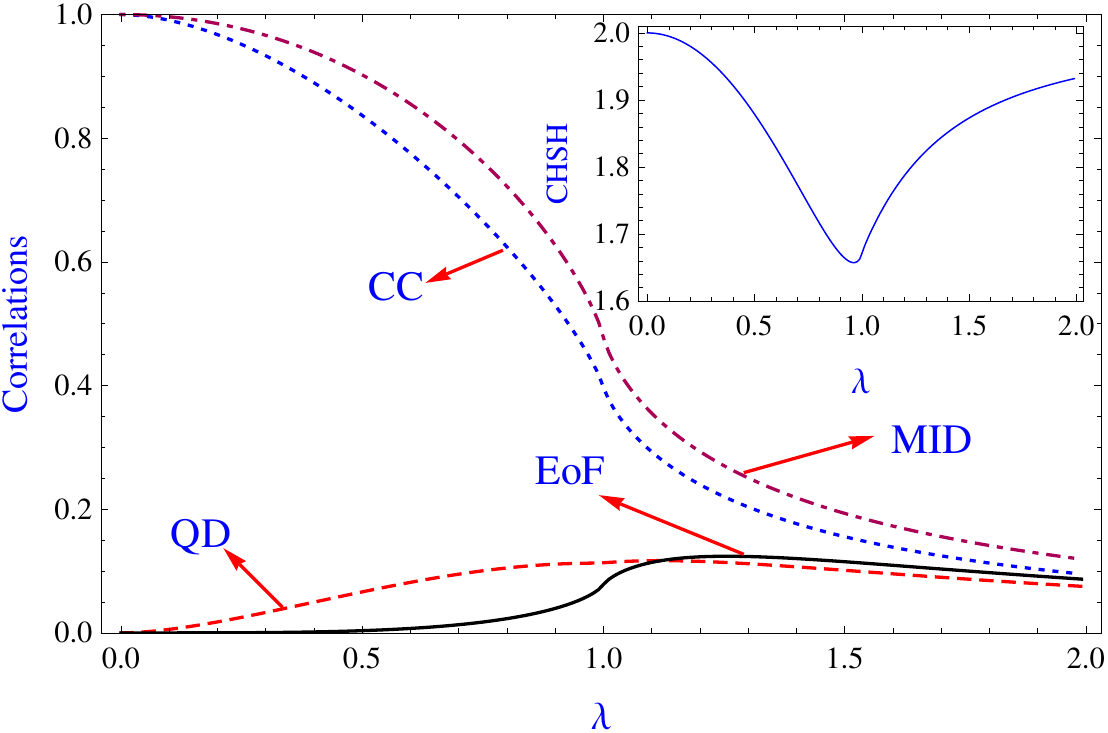}
\caption{(Color online) Correlation measures as a function of $\lambda$ for nearest-neighbor spins in the transverse field Ising model~($\gamma=1$ and $\alpha=0$) for a finite chain with $N=2001$ sites.}
\end{figure}

In Fig.~1, we display the correlation measures as a function of $\lambda$ for a chain with $N=2001$ sites in the transverse field Ising model. When the parameter $\lambda=0$, all spins are oriented in either the positive or negative z-direction and the ground state is doubly degenerate given by the density matrix:
\begin{eqnarray}
\rho=\frac{1}{2}\left|++\ldots+\right\rangle\left\langle ++\ldots+\right|+\frac{1}{2}\left|--\ldots-\right\rangle\left\langle --\ldots-\right|,
\end{eqnarray}
where $\left|\pm\right\rangle=(\left|\uparrow\right\rangle\pm\left|\downarrow\right\rangle)/\sqrt{2}$. After taking a partial trace of $\rho$ overall spins except the nearest neighbors, the two-point reduced density matrix in the standard basis would be
\begin{eqnarray}
\rho_{ii+1}=\left (\begin{array}{cccc} 0.25  & 0 & 0  & 0.25 \\ 0  & 0.25 & 0.25  & 0 \\ 0  & 0.25 & 0.25  & 0 \\ 0.25  & 0 & 0  & 0.25 \end{array} \right) \ ,
\end{eqnarray} 
which contains $EoF=QD=0$, but $MID=CC=1$ as can be seen from Fig.~1. On the other hand, for $\lambda\rightarrow\infty$ all spins are oriented along the positive z-direction and the ground state density matrix is equal to $\rho=\left|\uparrow\uparrow\ldots\uparrow\right\rangle\left\langle\uparrow\uparrow\ldots\uparrow\right|$ which is a product state that contains no type of correlations between any parties. The transition between the paramagnetic and ordered ferromagnetic phases occurs at $\lambda_c=1$.

As the magnitude of the external field increases from zero, QD and EoF increase and they attain their maximum value near the critical point~\cite{sarandy,dillenschneider}. On the other hand, CC and MID are monotonous functions of $\lambda$; the magnitudes of CC and MID decrease as $\lambda$ increases. Similar to EoF and QD, CC and MID increase when approaching the critical point from right for $\lambda>1$. The CHSH-Bell function, although below 2, reaches its minimum value in a region that is very close to $\lambda_c=1$~(see inset in Fig.~1). Fig.~1 also shows that $\lambda$-dependence of MID is, peculiarly, very close to that of CC and both of them are equal to $1$ for $\lambda=0$ where $EoF=QD=0$. One should note that although MID is easier to calculate than QD, it shows an inconsistency to reflect the quantum correlations compared to QD. It being a true quantum correlation measure as defined in Sec.~II, has been questioned, recently~\cite{gpa,wpm,sclmmp}. It is claimed that the lack of any optimization procedure over the projective measurements in the definition of MID makes it a weak correlation measure for some states~\cite{gpa,wpm,sclmmp}.
\begin{figure}[!ht]\centering
\includegraphics[width=5.9cm]{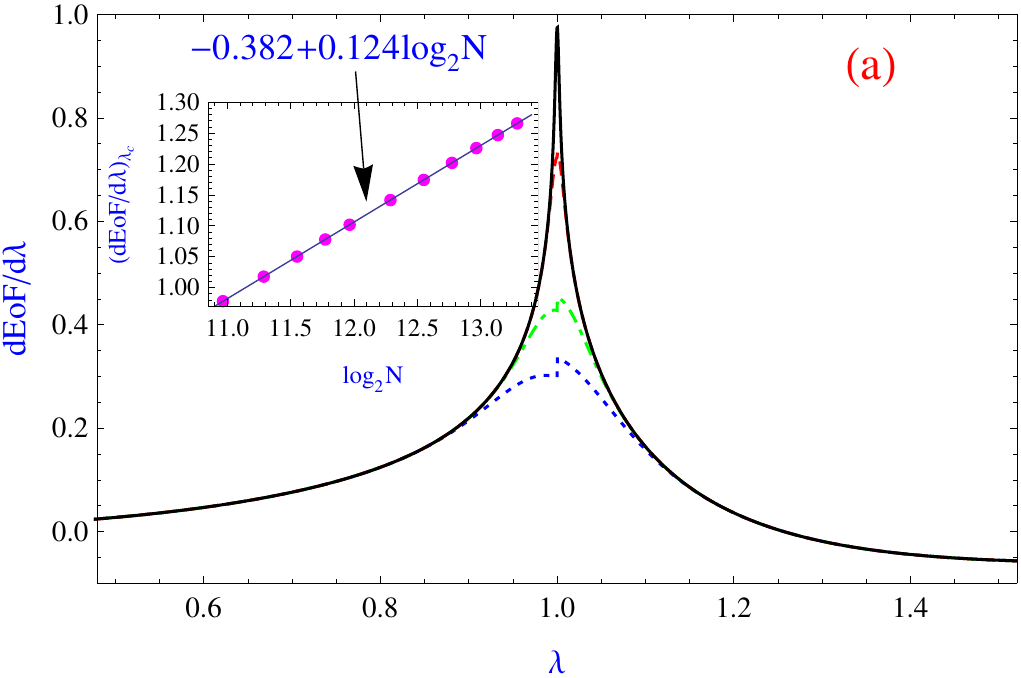}
\includegraphics[width=5.9cm]{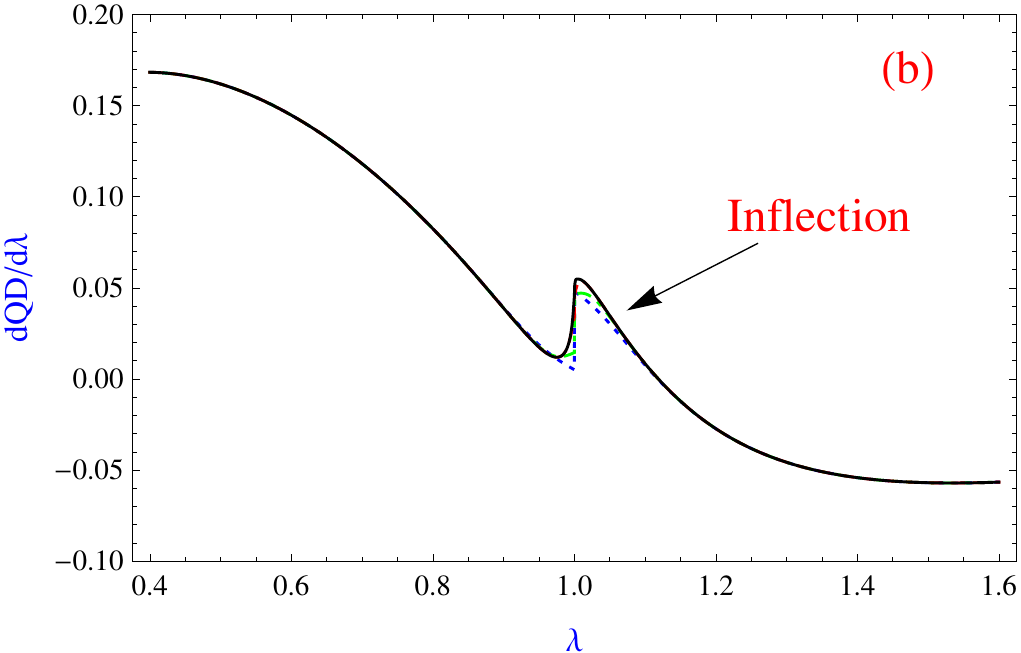}

\includegraphics[width=5.9cm]{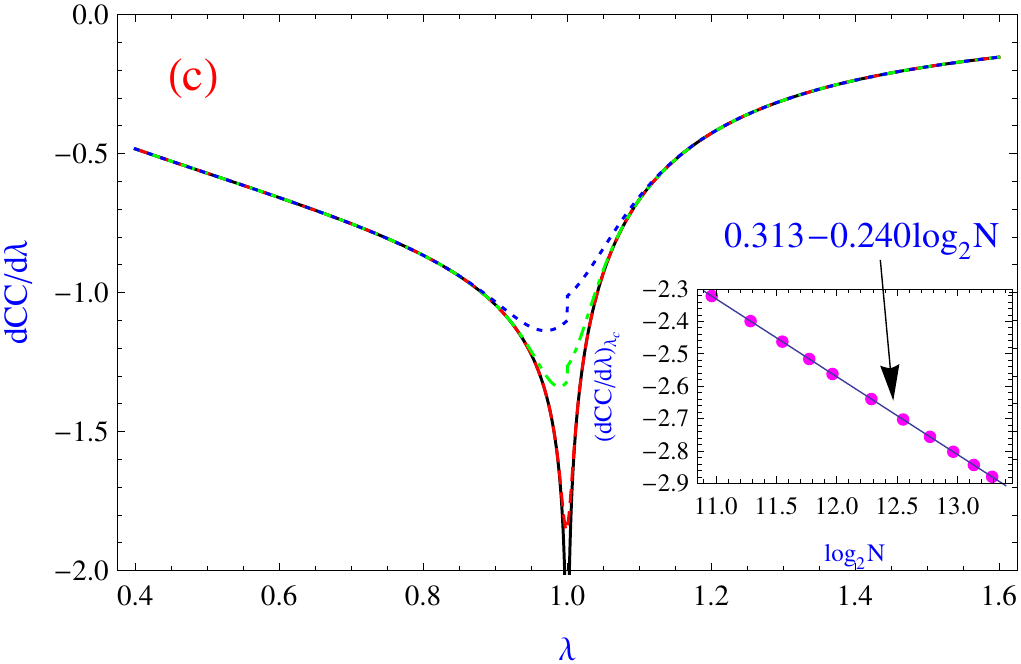}
\includegraphics[width=5.9cm]{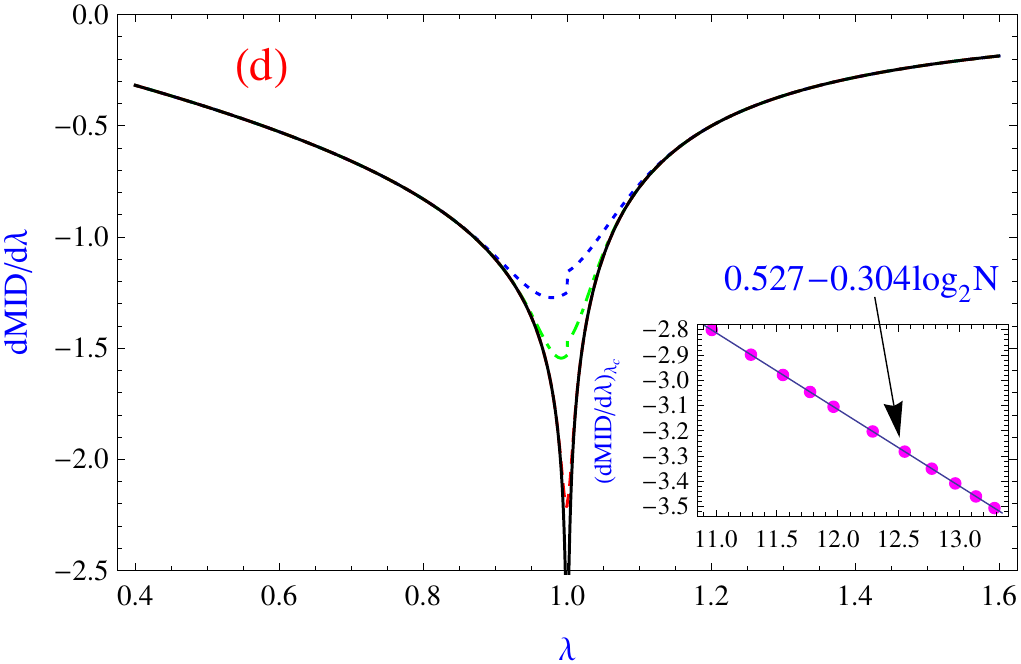}

\includegraphics[width=5.9cm]{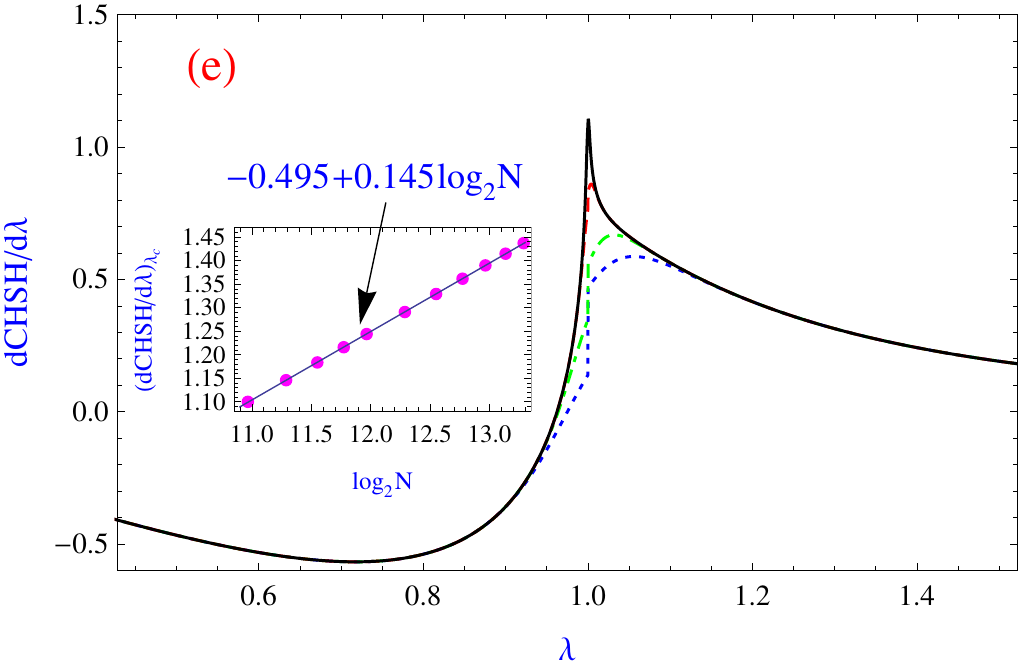}
\includegraphics[width=5.9cm]{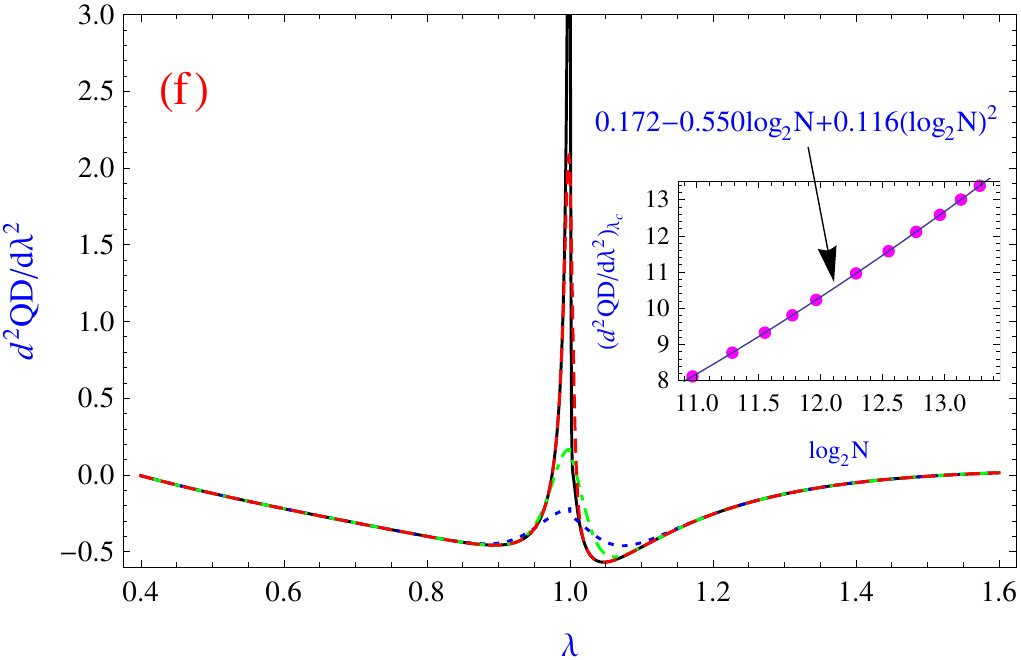}
\caption{(Color online) The first derivative of correlation measures~((a)-(e)) and the second derivative of QD~(f) for nearest-neighbor spins as a function of $\lambda$ in the transverse filed Ising chain~($\gamma=1$ and $\alpha=0$) for lattice sizes $N=51$~(blue-dotted), $N=101$~(green-dotdashed), $N=501$~(red-dashed) and $N=2001$~(black-solid). Insets: Finite size scaling behavior of the derivative(s) of the correlations at the critical point $\lambda_c=1$. Here filled circles are the numerical data, while the lines are the fitting curves fitted to $y=ax+b$ for (a)-(e) and to $y=ax^2+bx+c$ for (f).}
\end{figure}

The QPT at $\lambda_c=1$ is a second-order phase transition, so only the derivative of the correlation measures can signal the presence of the QPT as can be seen in Fig.~2. Indeed, the first derivatives of EoF~(Fig.~2(a)), CC~(Fig.~2(c)), MID~(Fig.~2(d)) and CHSH~(Fig.~2(d)) show pronounced minimums or maximums at some values of $\lambda$ which tend to the quantum critical point, $\lambda_c=1$, as the system size approaches infinity~($N\rightarrow\infty$). The insets in Fig.~2 show that these correlations exhibit a logarithmic divergence at $\lambda_c$. Peculiarly, the first derivative of QD behaves differently; it shows an inflection point around $\lambda_c$~(see Fig.~2(b)), while its second derivative shows a pronounced maximum which has a quadratic logarithmic divergence at $\lambda_c$~(see Fig.~2(f) and Ref.~\cite{sarandy}). It is remarkable to note that the behavior of QD is rather different from the other correlation measures considered in this paper.
\subsubsection{Transverse field XY model}
The second model that we will consider in this study is the transverse field XY model which can be obtained by setting $\alpha=0$ in Eq.~(\ref{xythamiltonian}):
\begin{eqnarray}\label{xyhamiltonian}
H_{XY}=-\sum_{j=1}^N\left\{\left(\frac{1+\gamma}{2}\right)\sigma_j^x\sigma_{j+1}^x+\left(\frac{1-\gamma}{2}\right)\sigma_j^y\sigma_{j+1}^y+\lambda\sigma_j^z\right\}.
\end{eqnarray}
The XY model has  interesting quantum phase transitions belonging to two different universality classes~\cite{mzpt}: {\bf (i)} For non-zero values of anisotropy, $\gamma$, the model exhibits a second-order quantum phase transition at the critical point $\lambda_c=1$. In the interval $0<|\gamma|\leq1$, this quantum phase transition has a universality class which is the same as that of the transverse field Ising model. {\bf (ii)} For $\lambda\in(0,1)$, the model exhibits again a second order phase transition, the so-called anisotropy transition, at the critical point $\gamma_c=0$. This critical point separates a ferromagnetic ordered along $x$-direction and the ferromagnetic ordered along the $y$-direction.
\begin{figure}[!ht]\centering
\includegraphics[width=5.9cm]{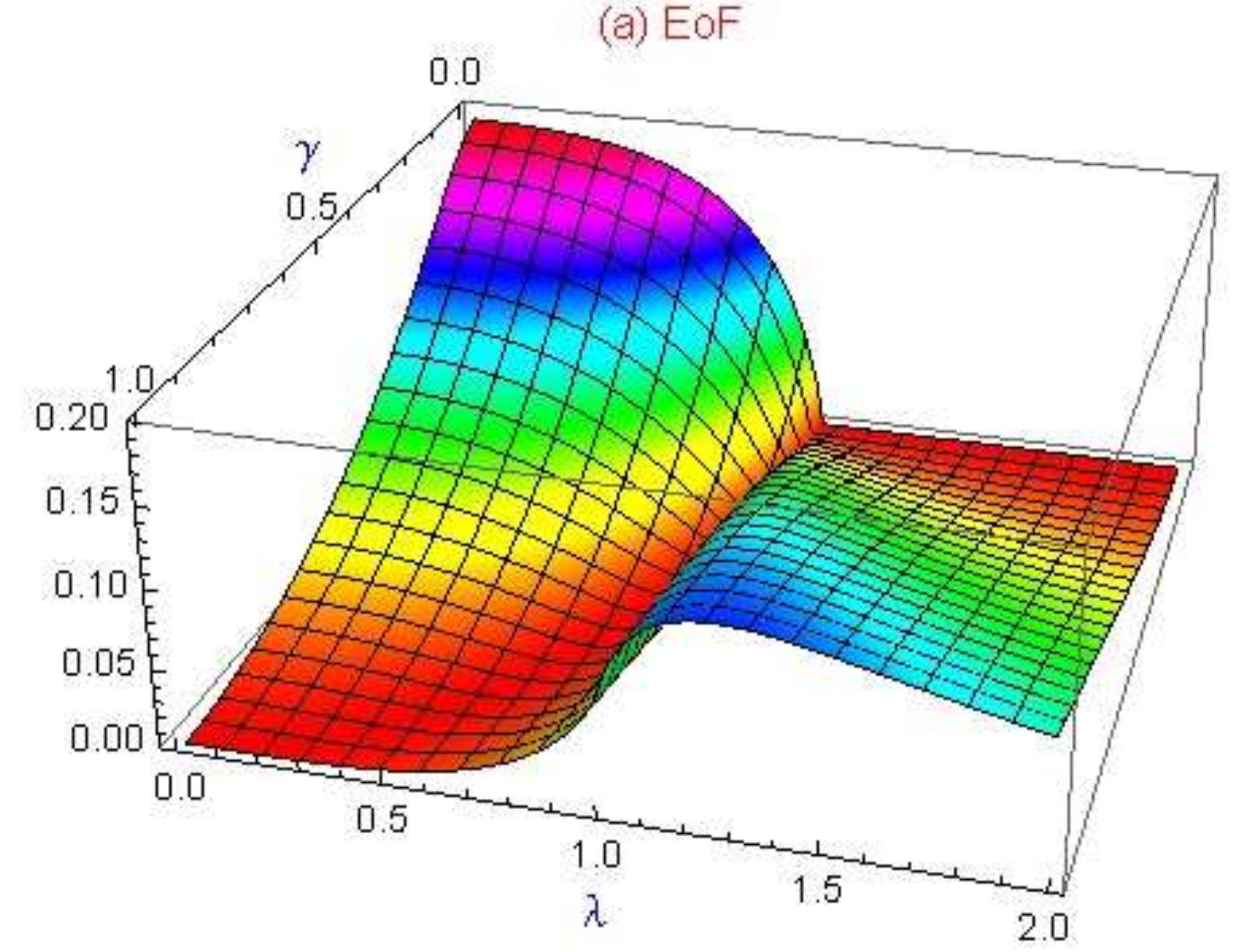}
\includegraphics[width=5.9cm]{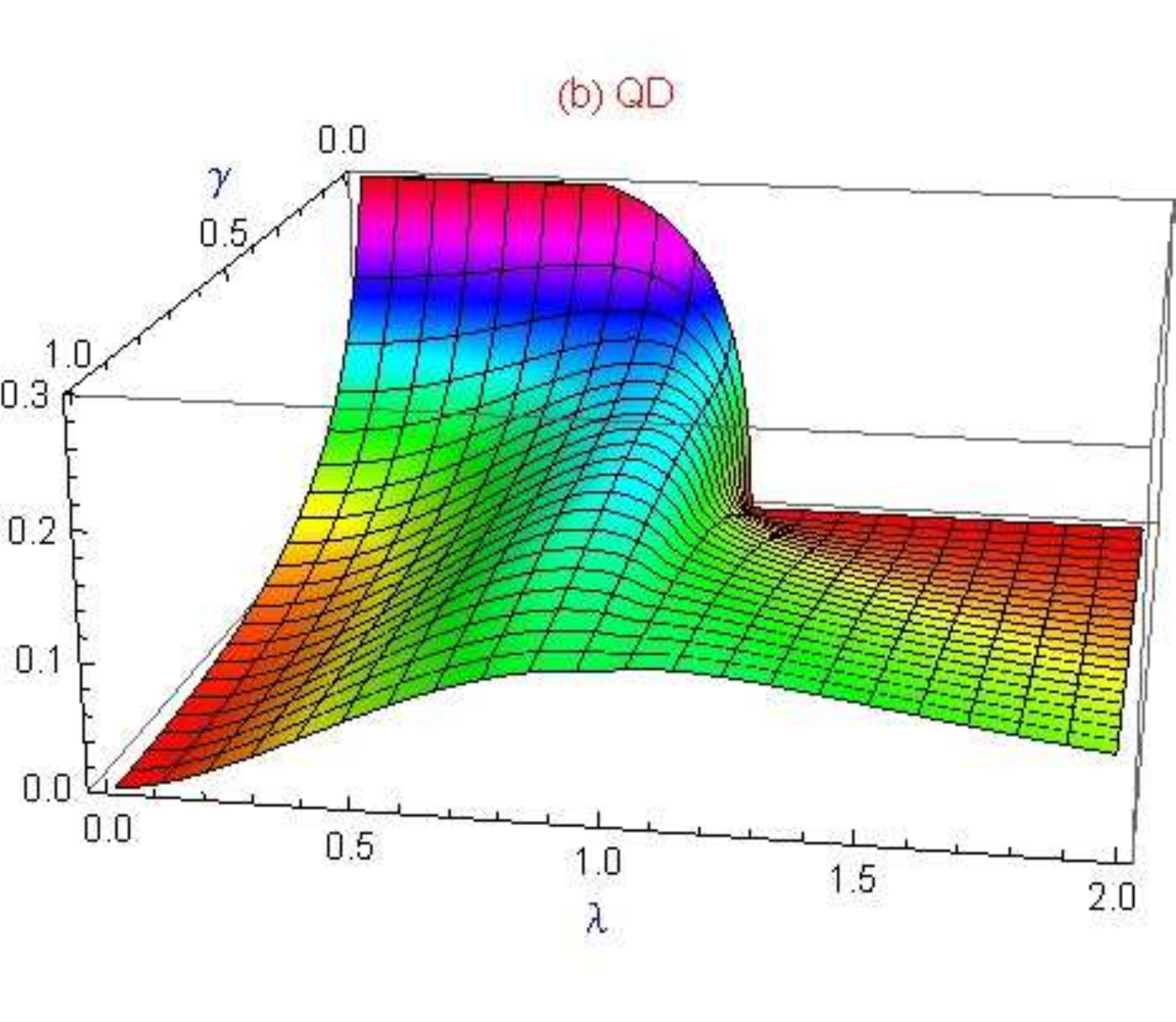}

\includegraphics[width=5.9cm]{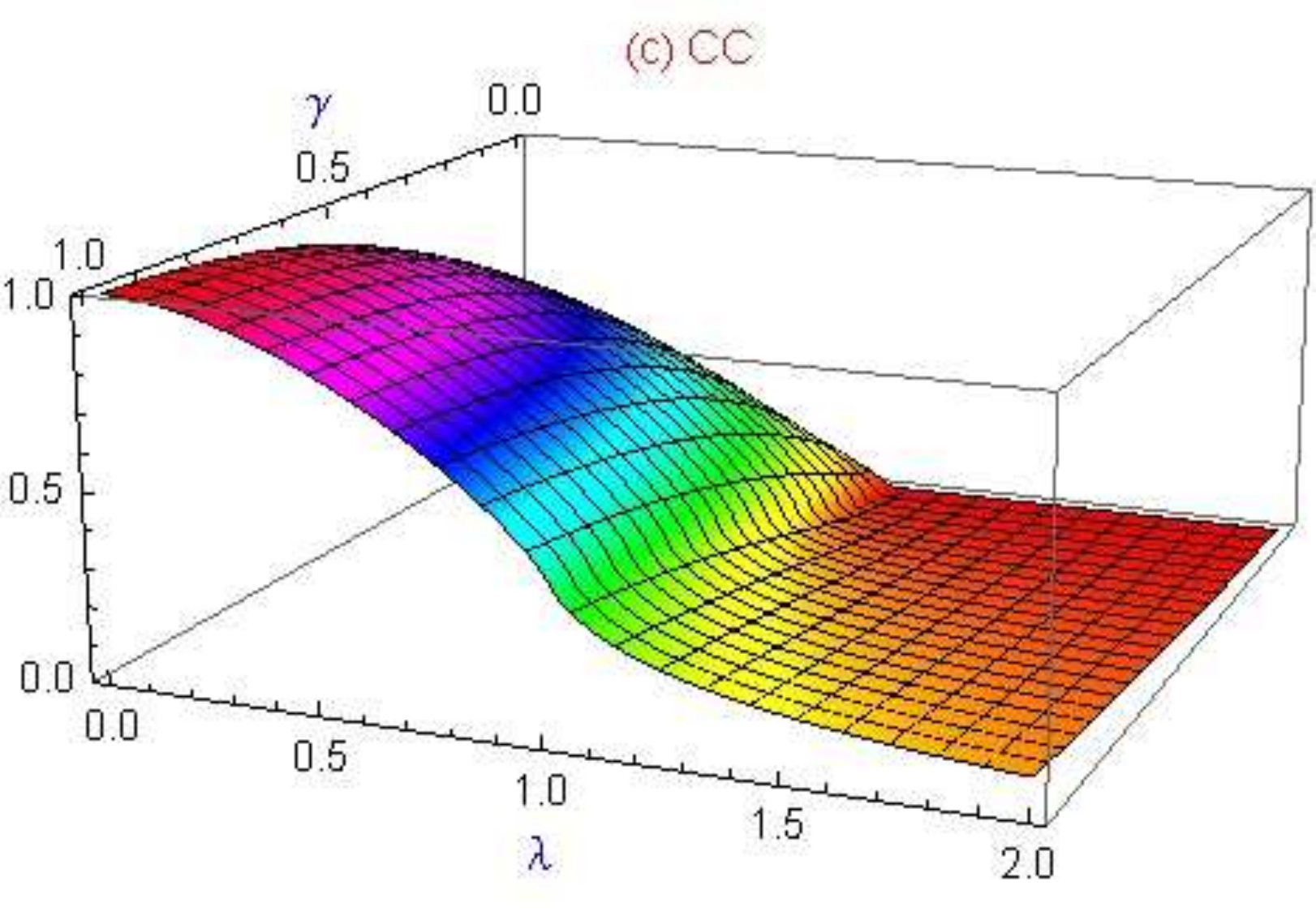}
\includegraphics[width=5.9cm]{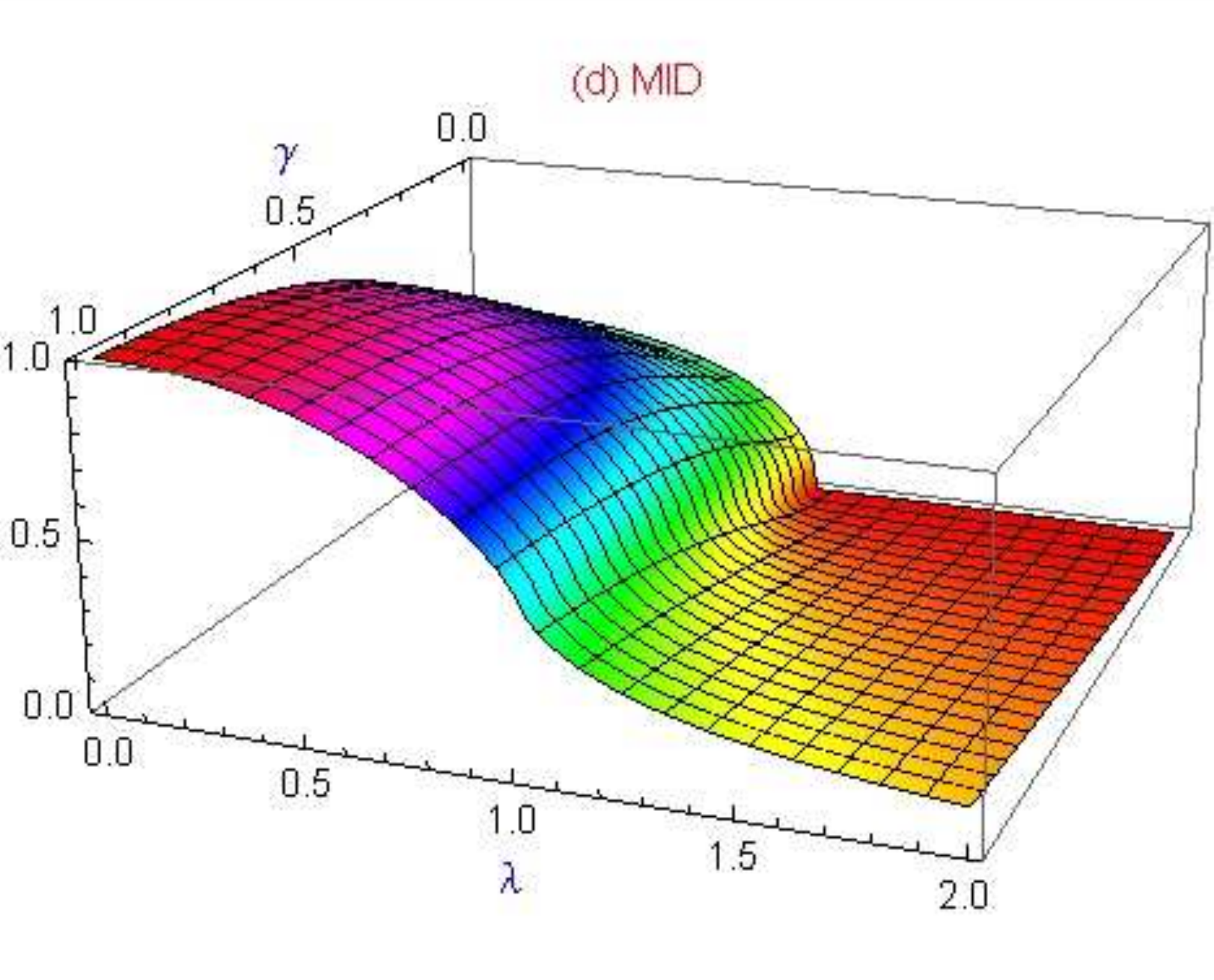}

\includegraphics[width=5.9cm]{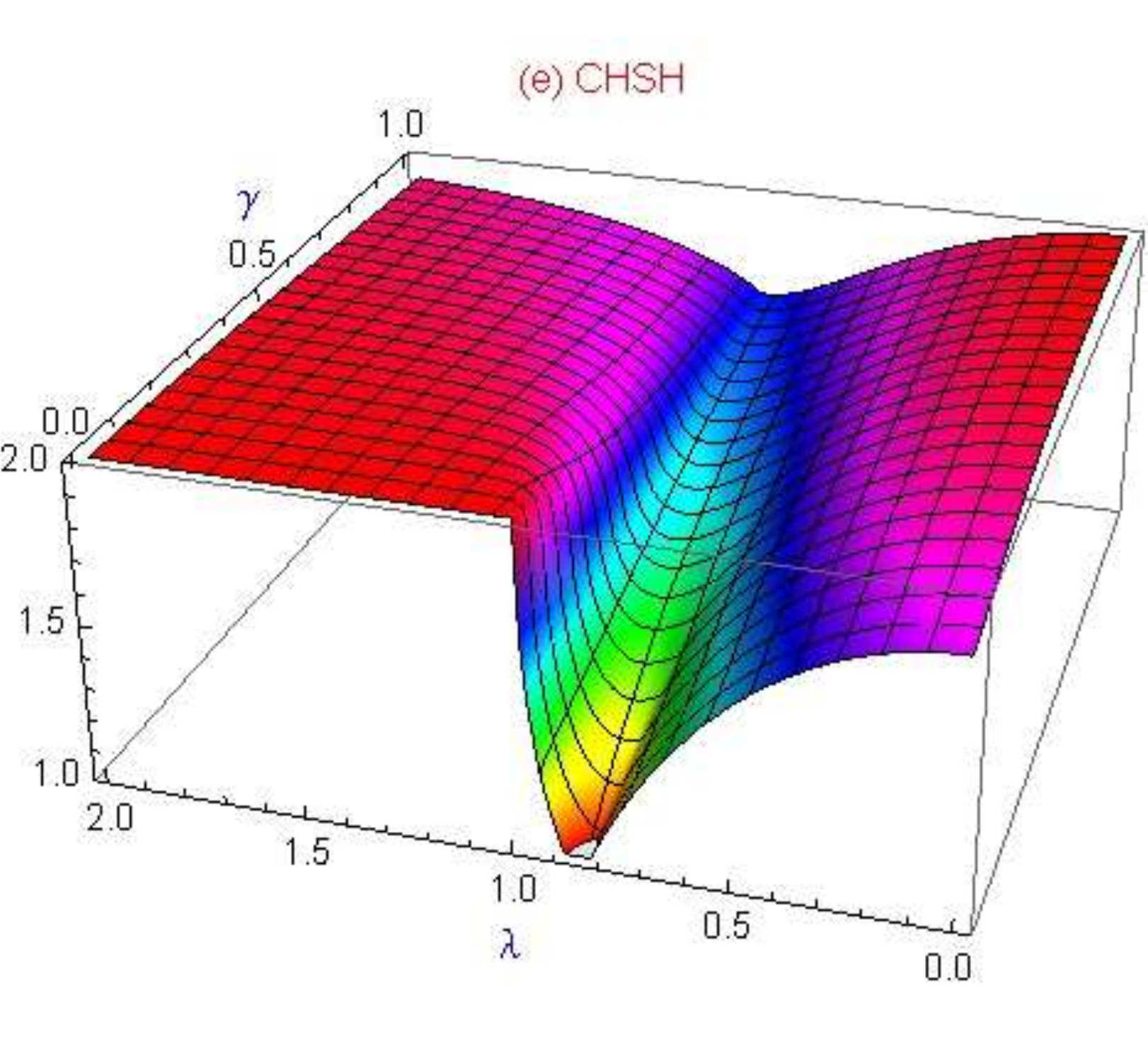}
\caption{(Color online) EoF~(a), QD~(b), CC~(c), MID~(d), CHSH~(e) in the XY spin chain~($\alpha=0$) with $N=2001$ sites as a function of $\gamma$ and $\lambda$.}
\end{figure}

In Fig.~3, we display EoF~(a), QD~(b), CC~(c), MID~(d) and CHSH function~(e) for nearest-neighbor spins in the XY spin chain with $N=2001$ as a function of $\gamma$ and $\lambda$ at zero temperature. All figures demonstrate that there is a pronounced difference in the correlation measures between the regions $0<\lambda<1$ and $1<\lambda<2$ where the QPT occurs at the critical line $\lambda_c=1$. When $\lambda<1$, EoF and QD decrease with $\gamma$. Contrarily, CC and MID increase. On the other hand, when $\lambda>1$, the correlations~(MID, QD, EoF and CC) increase with anisotropy. Actually, MID and CC are increasing functions of $\gamma$ in the considered range of $\lambda$. The CHSH-Bell function is always below 2, i.e., the Bell inequality is not violated, and its dependence on $\gamma$ is weak for the regions far from the critical point $\lambda_c=1$. Near the critical point, CHSH attains its minimum. The minimum decreases when anisotropy decreases and becomes more sharp when $\gamma\rightarrow0$~($XX$ model).
\begin{figure}[!ht]\centering
\includegraphics[width=5.9cm]{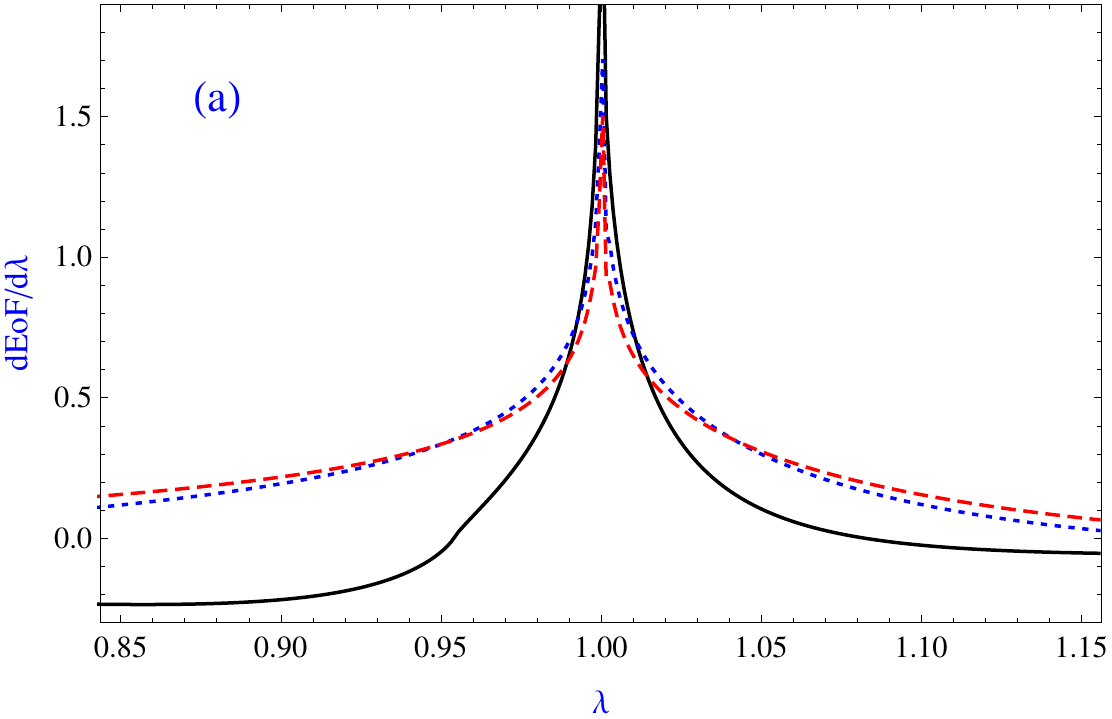}
\includegraphics[width=5.9cm]{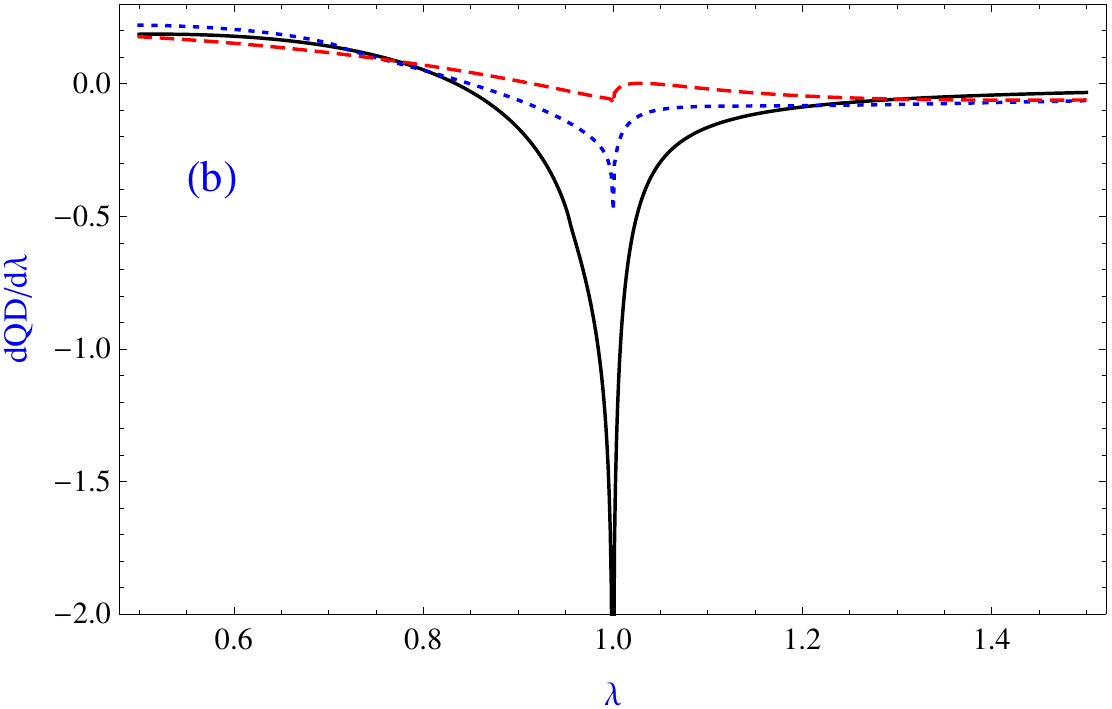}

\includegraphics[width=5.9cm]{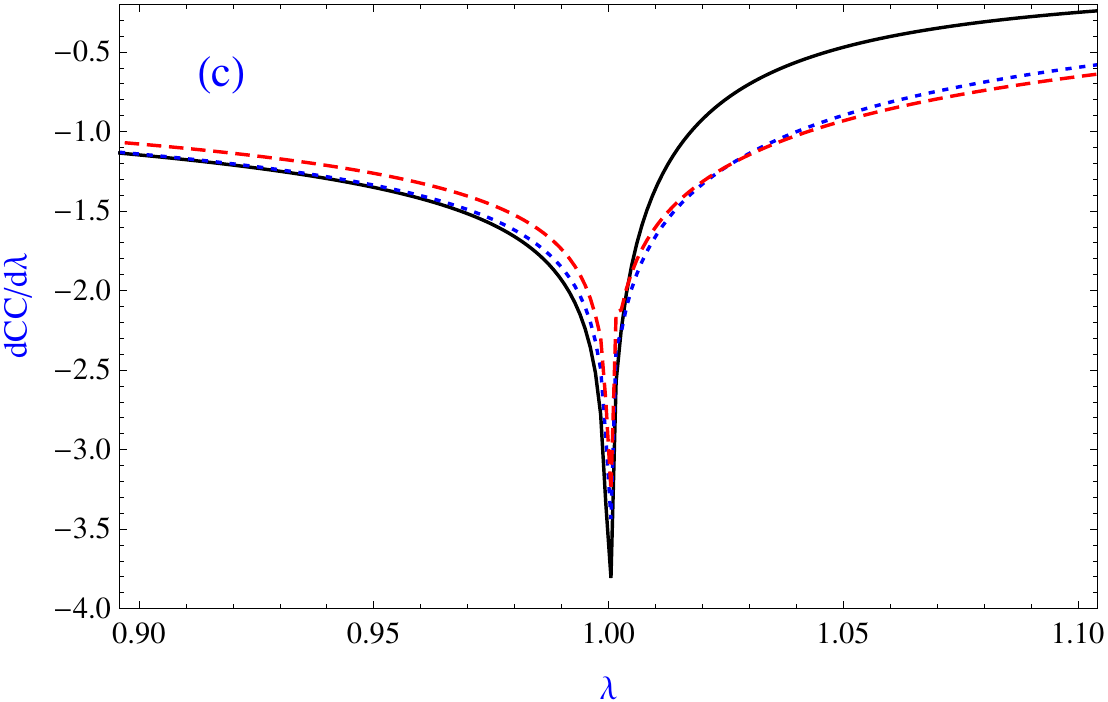}
\includegraphics[width=5.9cm]{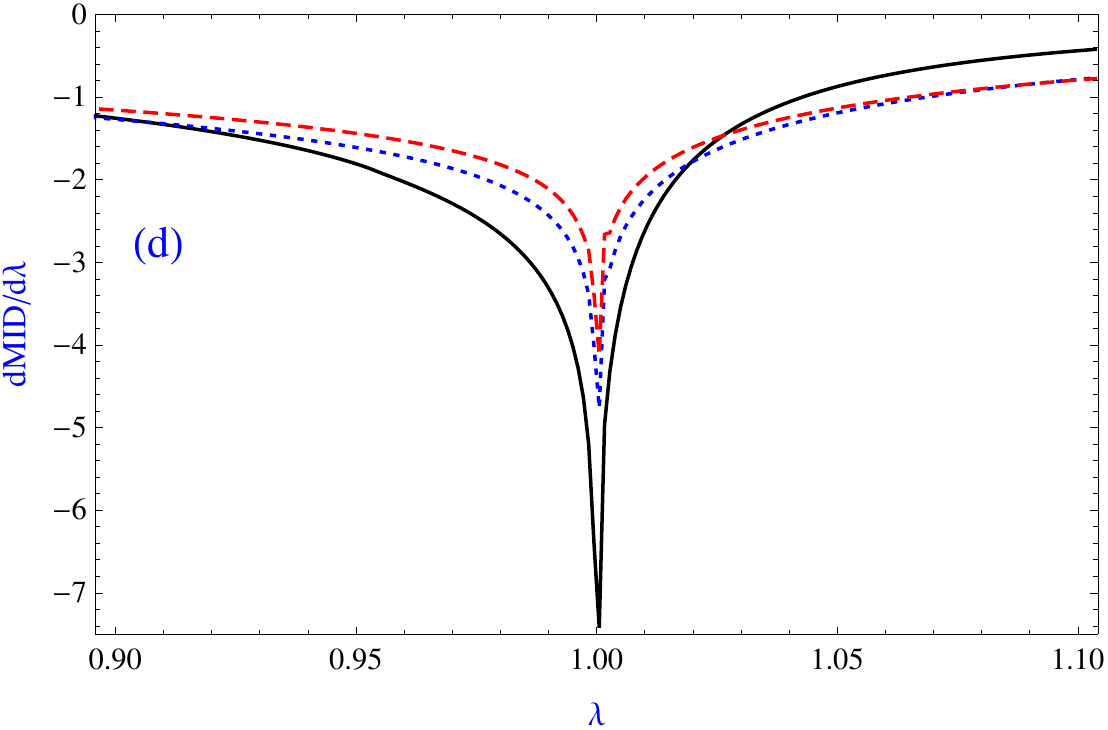}

\includegraphics[width=5.9cm]{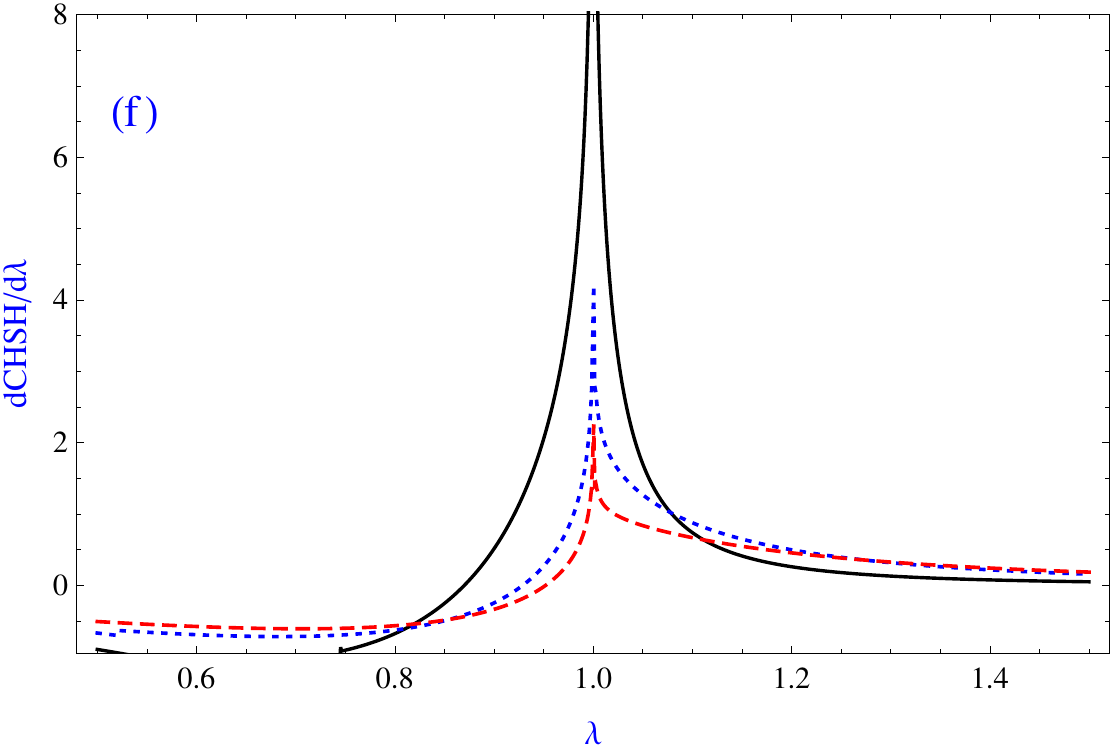}
\caption{(Color online) The first $\lambda$-derivative of correlations, EoF~(a), QD~(b), CC~(c), MID~(d) and CHSH~(e) as a function of $\lambda$ in the anisotropic XY model~($\alpha=0$) for different values of anisotropies $\gamma=0.3$~(black-solid), $\gamma=0.7$~(blue-dotted) and $\gamma=0.9$~(red-dashed) and $N=2001$.}
\end{figure}

As mentioned above, the XY model undergoes a second-order Ising like quantum phase transition at $\lambda_c=1$. Thus, in Fig.~4, we display the first derivative of correlations as a function of $\lambda$ for different values of $\gamma$ and $N=2001$. As depicted in these subfigures, the first derivative of all correlation measures are divergent at $\lambda_c=1$, signaling the presence of the QPT. It seems that the decrease in $\gamma$ sharpens the singularities to some degree. One should note that the derivative of QD has a different character for the XY model compared to that for the Ising model discussed above. For a finite chain Ising model~($\gamma=1$ and $N=2001$), the first derivative of QD is found to exhibit an inflection point around $\lambda_c$, while its second derivative is divergent at $\lambda_c$~(see Figs.~2(b) and~2(f)). As shown in Fig.~4(b), for the anisotropic XY model, its first derivative is already divergent at $\lambda_c$.
\begin{figure}[!ht]\centering
\includegraphics[width=7.9cm]{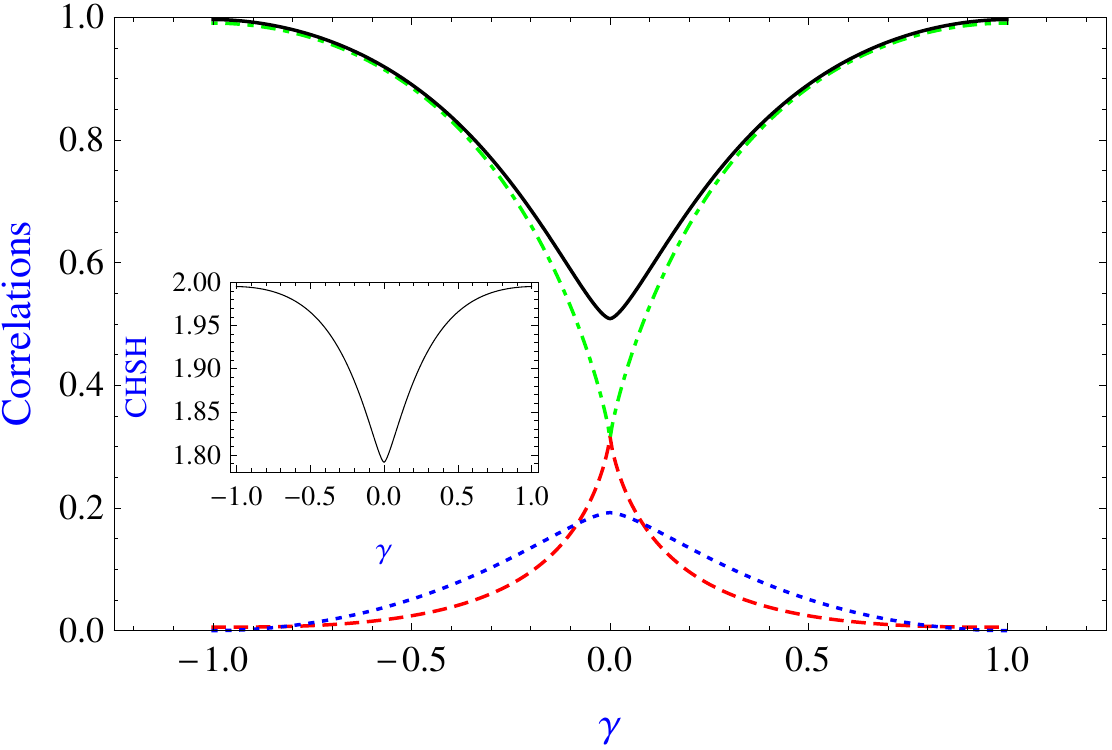}
\caption{(Color online) QD~(red-dashed), EoF~(blue-dotted), CC~(green-dotdashed), MID~(black-solid) and CHSH-Bell function~(inset) as a function of $\gamma$ for the anisotropic XY model~($\alpha=0$) with $N=2001$ and $\lambda=0.1$.}
\end{figure}

In Fig.~5, we plot EoF, QD, MID, CC and CHSH-Bell function~(inset) as a function of anisotropy $\gamma$ for $N=2001$ and $\lambda=0.1$. All correlation measures are symmetric according to $\gamma\rightarrow-\gamma$. EoF and QD are maximum at $\gamma_c$, while MID, CC and CHSH function reach their minimums. It is safe to say that all the considered correlation measures precisely detect the anisotropy phase transition by reaching their maximums or minimums. Especially, the behaviors of QD and CC are more drastic, showing extremum at $\gamma_c$ with a cusplike behavior arising from the minimization procedure over the complete set of projectors in their definitions. In fact, the first $\gamma$ derivative of EoF, MID and CHSH show an inflection around the critical point $\gamma_c=0$, while the first derivatives of QD and CC are discontinuous at $\gamma_c$. On the other hand, second derivatives of all correlations diverge at the anisotropy transition point~(not shown here).
\subsubsection{XYT model}
Now, we consider the general Hamiltonian, Eq.~(\ref{xythamiltonian}), and analyze the correlation measures in the ground state of the XYT model. The model has a rich quantum phase diagram~(see Refs.~\cite{lzxt,ýtgj}). Differently from the transverse field XY model, the QPTs are caused not only by the external field and anisotropy, but also by the three spin interactions. Indeed, the competition between $\lambda$, $\gamma$ and $\alpha$ may induce QPTs belonging to different universality classes, for example, ferromagnetic, chiral and different spin liquid phases~\cite{lzxt,ýtgj}.
\begin{figure}[!ht]\centering
\includegraphics[width=6.9cm]{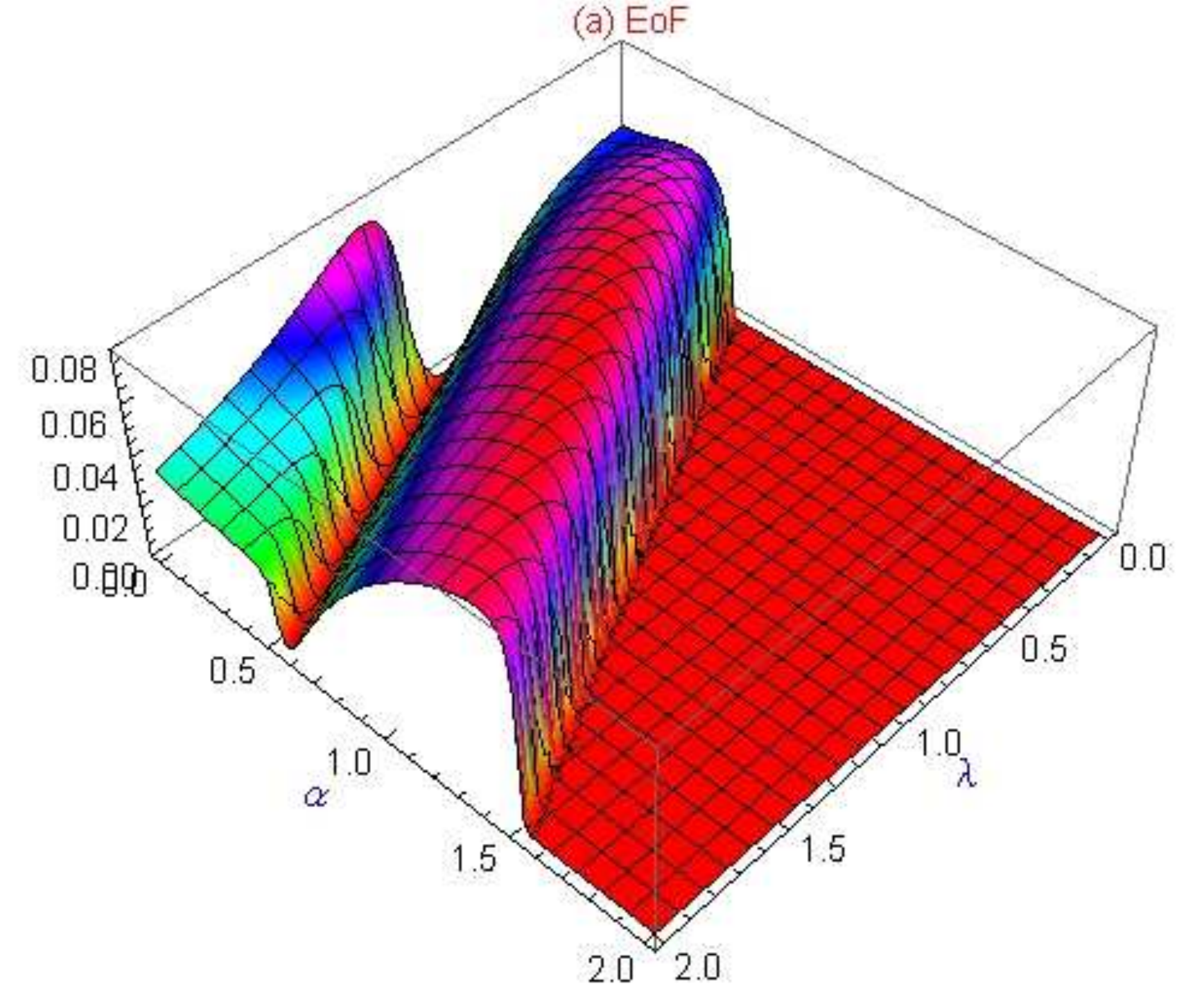}
\includegraphics[width=6.9cm]{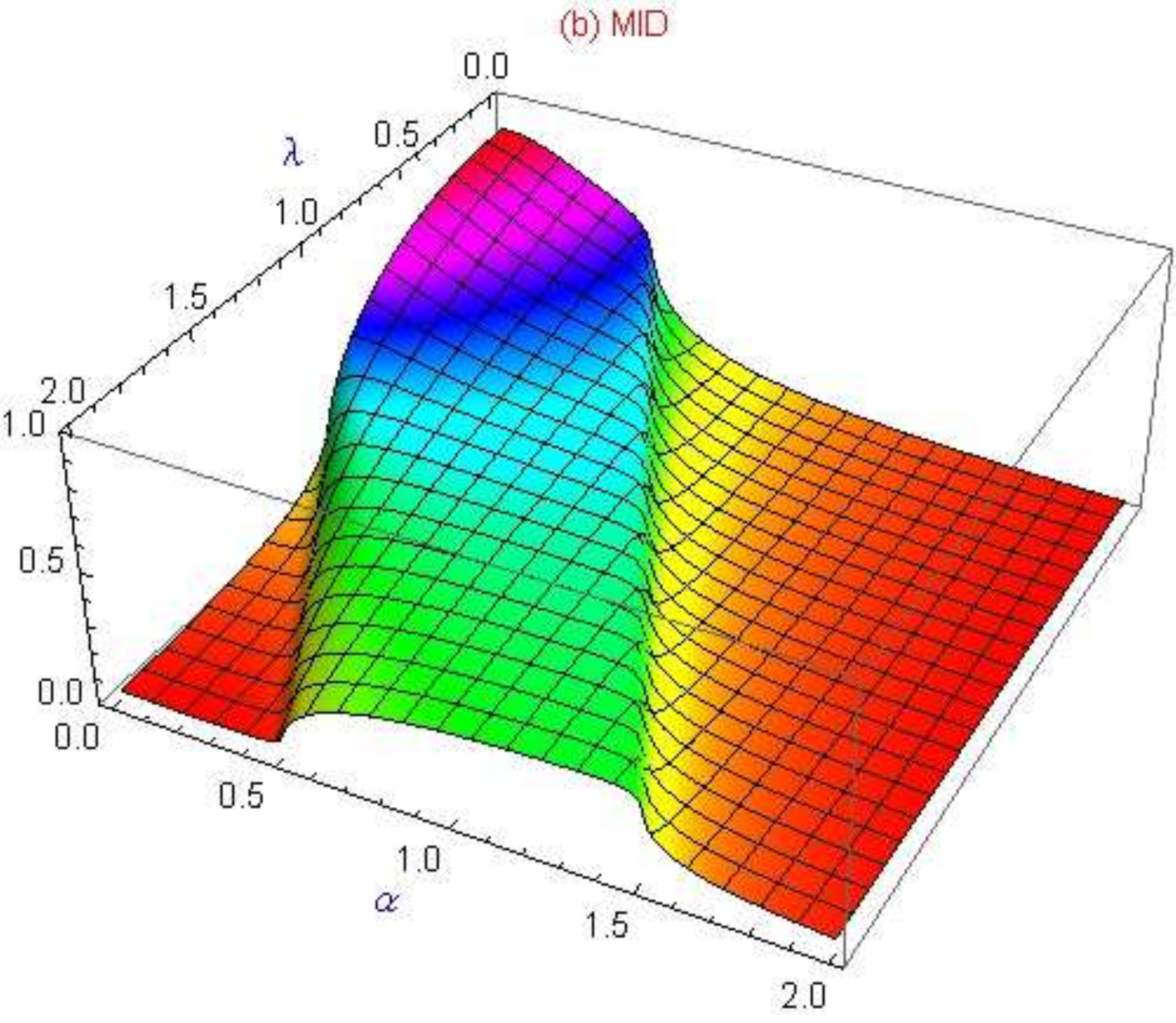}
\caption{\label{figxyt1} EoF~(a) and MID~(b) as a function of $\lambda$ and $\alpha$ at $\gamma=0.5$ and $N=2001$ in the XYT model.}
\end{figure}

We start the discussion of the correlation-phase diagram of XYT model with an investigation of the effect of three-spin interaction strength and the external magnetic field while the anisotropy is held constant. In Fig.~\ref{figxyt1}, we display EoF and MID as functions of $\lambda$ and $\alpha$ at $\gamma=0.5$ and $N=2001$ in the XYT model. The dramatic changes in EoF and MID show the phase transitions in the model~(similar for other considered correlation measures, thus we have not displayed them here). The region between $\alpha\in[0,0.5)$ and $\lambda\in(0.9,2.0)$ indicates the spin-saturated phase~(ferromagnetic), while the regions between $\alpha\in(0.5,1.5)$ and  $\lambda\in[0,2]$ as well as  $\alpha\in(1.5,2]$ and  $\lambda\in[0,2]$ are the two kinds of incommensurate phases with weak spontaneous magnetization~(see Refs.~\cite{ylhl,wcjl,zvyagin} for details). 
\begin{figure}[!ht]\centering
\includegraphics[width=5.9cm]{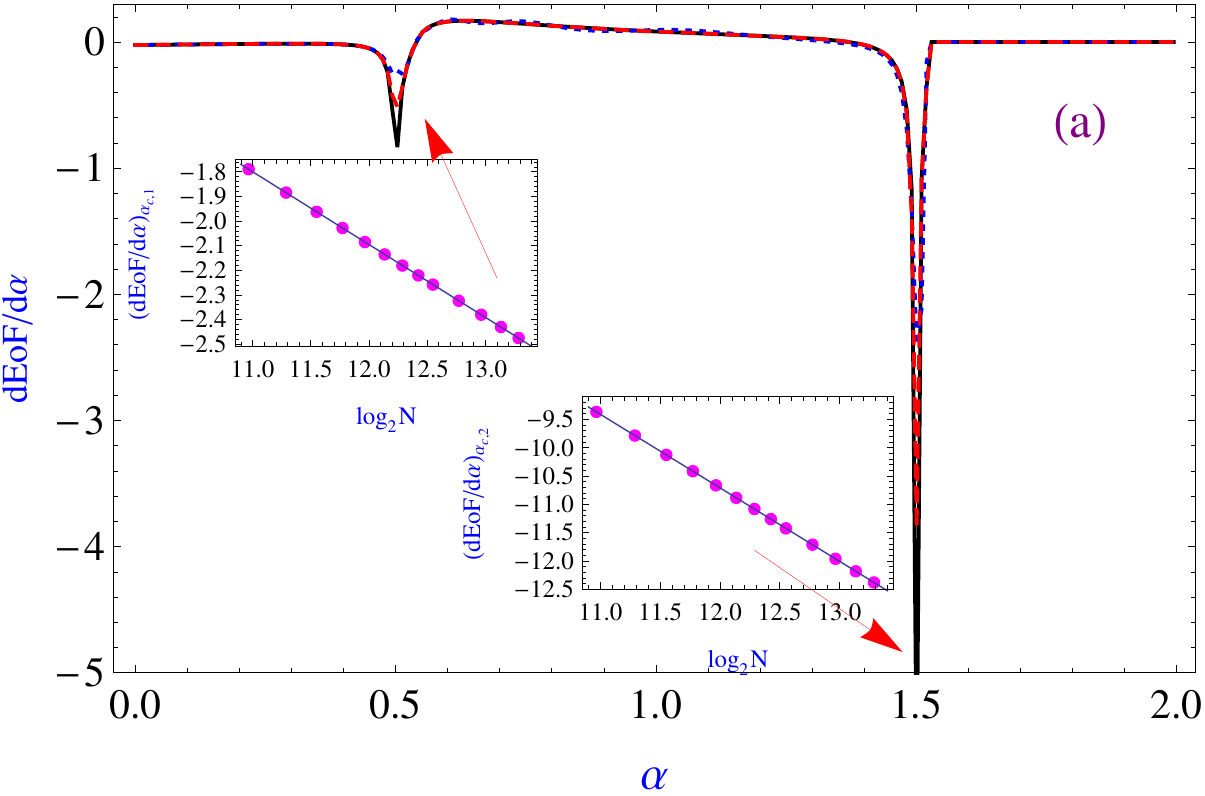}
\includegraphics[width=5.9cm]{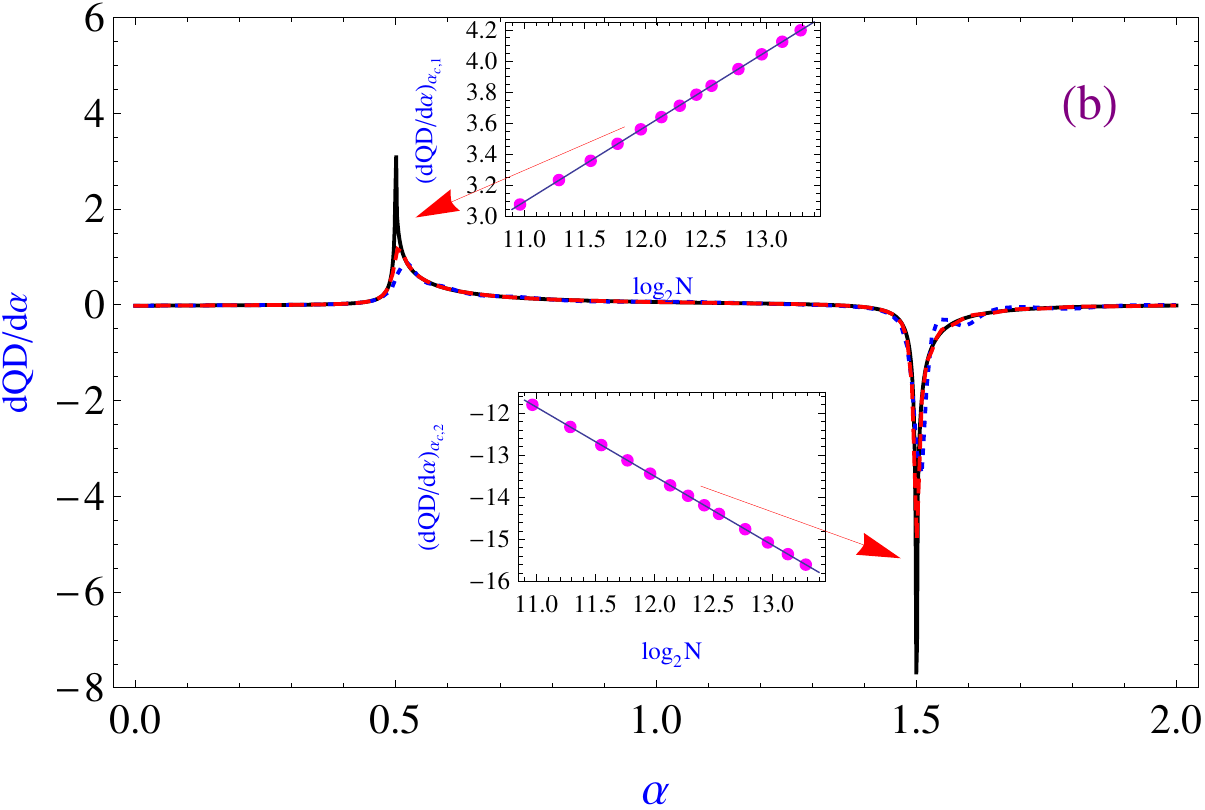}

\includegraphics[width=5.9cm]{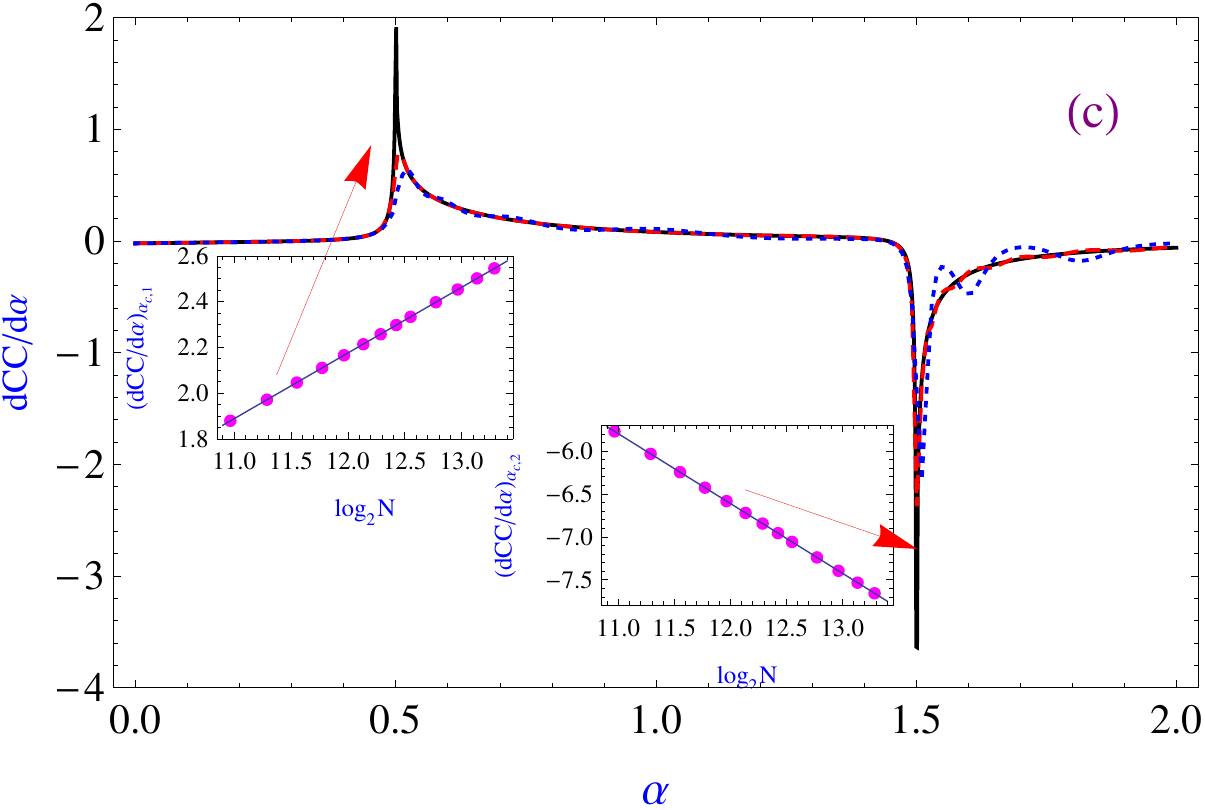}
\includegraphics[width=5.9cm]{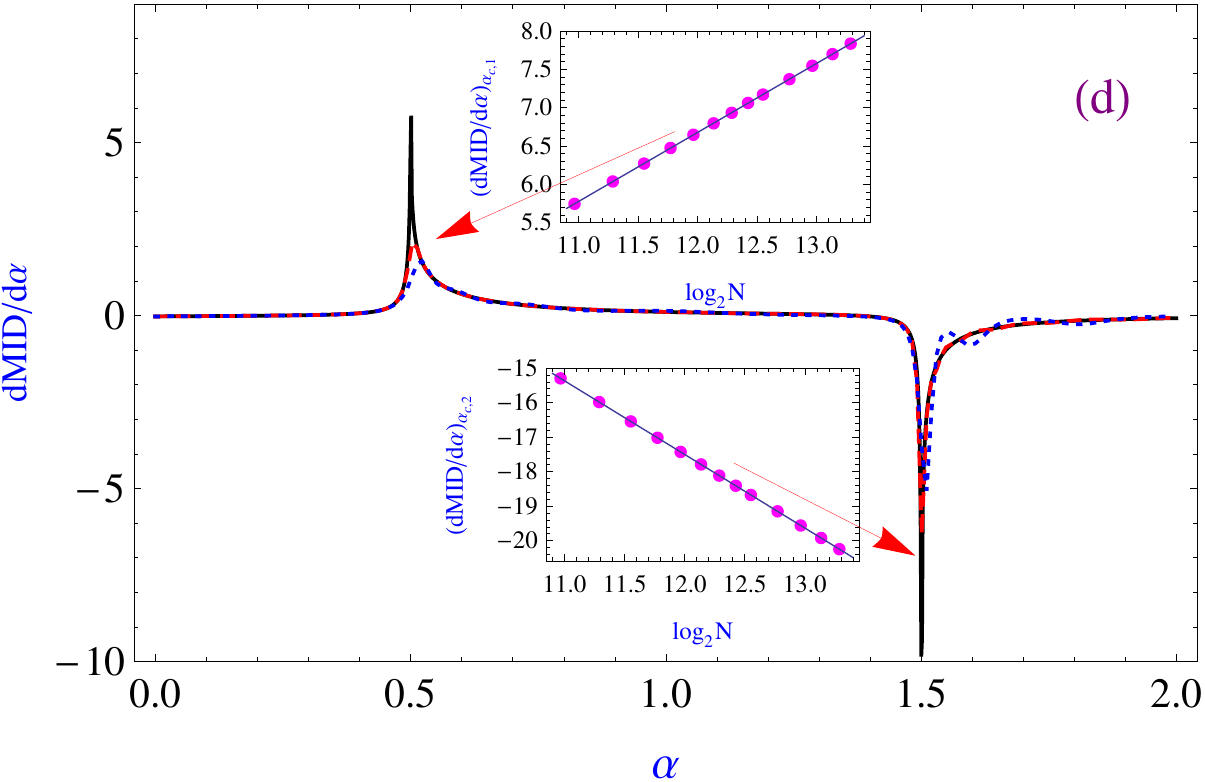}

\includegraphics[width=5.9cm]{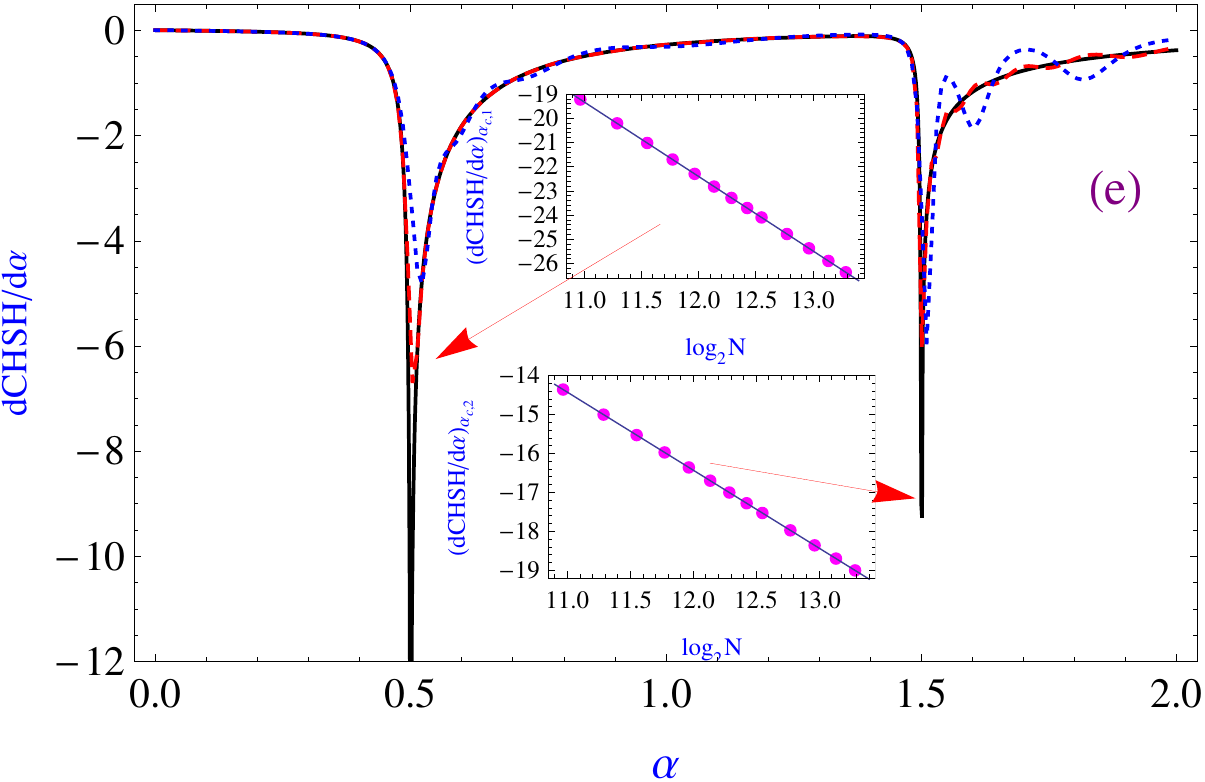}
\caption{\label{figxyt2}(Color online) The $\alpha$-derivative of correlation measures as a function of $\alpha$ at $\lambda=2$ and $\gamma=0.5$ in the XYT chain with $N=51$~(blue-dotted), $N=101$~(red-dashed) and $N=2001$~(black). Insets show the finite size scaling behavior of correlation measures at the QPTs $\alpha_{c,1}=0.5$ and $\alpha_{c,2}=1.5$ where the filled circles are numerical data, while the lines are the fitting curves fitted to $y=ax+b$.}
\end{figure}

All transitions are second-order, thus one can expect that the derivative of correlation measures could signal these QPTs. In Fig.~\ref{figxyt2}, the $\alpha$-derivatives of correlation measures as a function of $\alpha$ are displayed at $\lambda=2.0$ and $\gamma=0.5$ in the XYT model with different lattice sizes. The first derivatives of all correlation measures diverge at $\alpha_{c,1}=0.5$ and $\alpha_{c,2}=1.5$ showing the presence of QPTs. These extremum points indicate that there exists three different phases as a function of $\alpha$ in the region $\alpha\in[0,2]$ for the XYT model. Note that the first derivatives of all the correlations exhibit a logarithmic divergence at $\alpha_{c,1}$ and $\alpha_{c,2}$ as $N$ approaches infinity~(see insets in Fig.~\ref{figxyt2}).
\begin{figure}[!ht]\centering
\includegraphics[width=7.1cm]{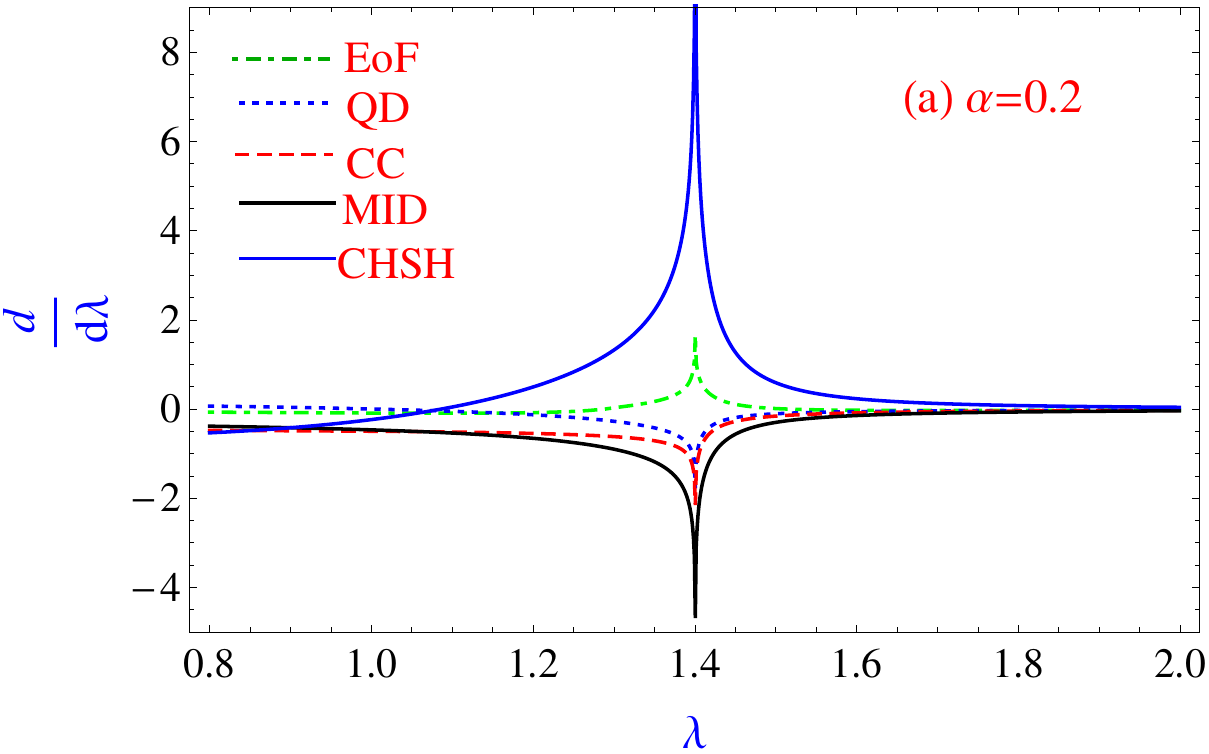}
\includegraphics[width=7.1cm]{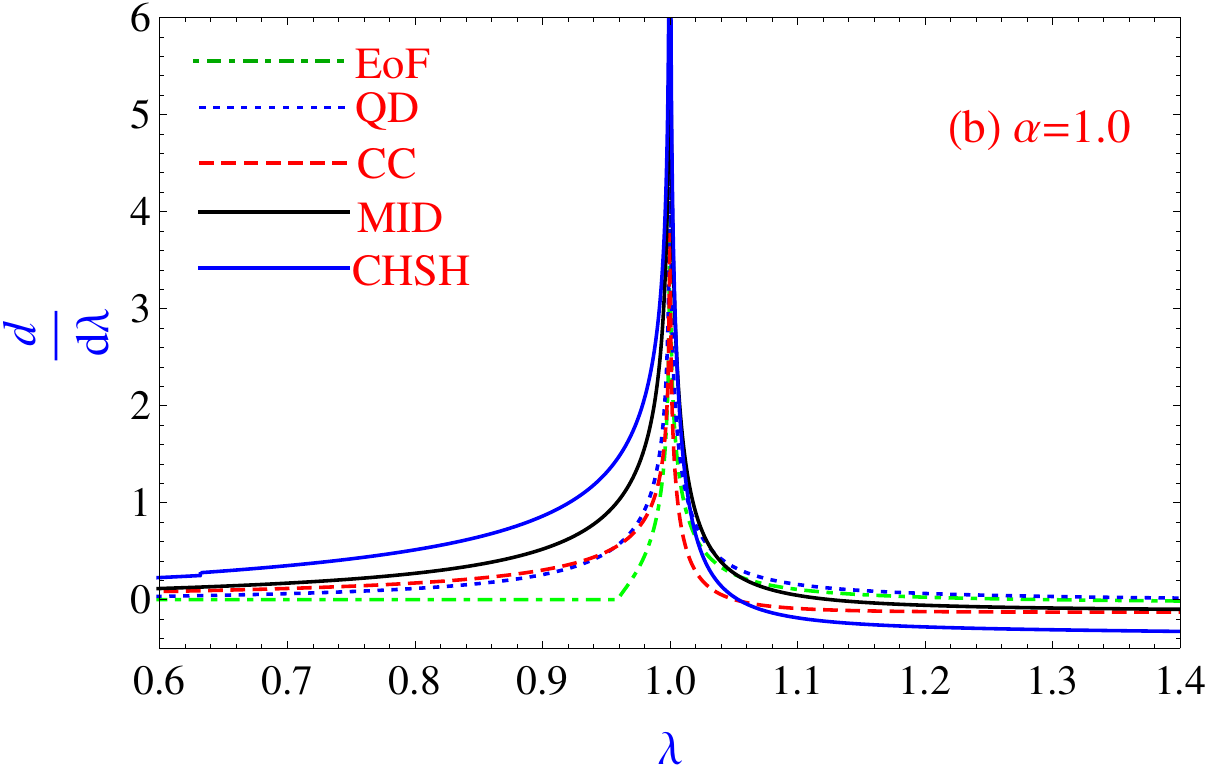}
\caption{\label{figxyt3}(Color online) First $\lambda$-derivative of correlation measures as  a function of $\lambda$ in the XYT model with $N=2001$, $\gamma=0.5$ and $\alpha=0.2$~(a) and $\alpha=1.0$~(b).}
\end{figure}

As shown in Fig.~\ref{figxyt1}, for a fix $\alpha$, the external magnetic field can also lead to QPTs in the XYT model. For instance, for $\alpha=0.2$ and $\alpha=1.0$, there are second-order quantum phase transitions at $\lambda_c=1.4$ and $\lambda_c=1.0$, respectively. The divergent behavior of the first derivative of correlations with respect to $\lambda$ signals these phase transitions as shown in Figs.~\ref{figxyt3}(a) and~\ref{figxyt3}(b). They diverge logarithmically as the system size increases, which is similar to the QPTs in the line of $\alpha$~(we have not displayed them here). In Ref.~\cite{ylhl}, the authors studied QPTs in the XYT model and they only focused on the critical point $\alpha_c=0.5$ for $\gamma=0.5$ and $\lambda=0$. It was shown that the first $\alpha$-derivative of QD and CC are capable of detecting the QPT. Different from Ref.~\cite{ylhl}, we have shown that QD and CC as well as MID, CHSH and EoF can signal all the phase transitions present in the XYT model for $\gamma=0.5$ and $\lambda>0$.
\subsection{XXZ model}\label{sec3}
The next model that we tackle in this subsection is the one dimensional anisotropic Heisenberg XXZ chain given by the Hamiltonian~\cite{dillenschneider}:
\begin{eqnarray}\label{xxzhamiltonian}
H_{XXZ}=\sum_{i=1}^NS_i^xS_{i+1}^x+S_i^yS_{i+1}^y+\Delta S_i^zS_{i+1}^z,
\end{eqnarray}
where $\Delta$ is the anisotropy parameter, $S_j^{\mu}=\sigma_j^{\mu}/2$ and $N$ is the number of particles in the chain. The model has two critical points that separates three different phases: {\bf (i)} At $\Delta_c=-1$, there is a first order QPT that separate the ferromagnetic phase, where all the spins are pointing in the same direction for $\Delta<-1$, and the gapless phase, where the correlation decays polynomially for $-1<\Delta<1$. {\bf (ii)} At $\Delta_c=1$, there is an infinite order QPT that separates the gapless phase and the anti-ferromagnetic phase for $\Delta>1$.

The Hamiltonian~(\ref{xxzhamiltonian})  has a spin-flip symmetry that leads to zero magnetization $\left\langle\sigma^z\right\rangle=0$ and it is symmetric with respect to the rotation in the $xy$ plane which leads to $\left\langle\sigma_i^x\sigma_j^x\right\rangle=\left\langle\sigma_i^y\sigma_j^y\right\rangle$.  Due to these symmetries, the nearest neighbor reduced density matrix would be in an X-state form as in Eq.~(\ref{xstate}) with matrix elements
\begin{eqnarray}
u_{11}=u_{44}&=&\frac{1+\left\langle\sigma_i^z\sigma_{i+1}^z\right\rangle}{4},\nonumber\\
u_{22}=u_{33}&=&\frac{1-\left\langle\sigma_i^z\sigma_{i+1}^z\right\rangle}{4},\nonumber\\
u_{23}=u_{32}&=&\frac{\left\langle\sigma_i^x\sigma_{i+1}^x\right\rangle}{2}, \quad u_{14}=0.
\end{eqnarray}

The XXZ model cannot be diagonalized, but the two point correlation functions at zero temperature, $T=0$, and in the thermodynamic limit, $N\rightarrow\infty$, can be obtained by using the Bethe ansatz technique~\cite{msmt}. The spin-spin correlation functions between nearest-neighbor spin sites
for $-1<\Delta< 1$ are given by~\cite{ksts}:
\begin{eqnarray}
\langle\sigma_i^z\sigma_{i+1}^z\rangle &=&1-\frac{2}{\pi^2}\int_{-\infty}^{\infty}\frac{dx}{\sinh x}\frac{x\cosh x}{\cosh^2(\Phi x)}+\frac{2\cot(\pi\Phi)}{\pi}\int_{-\infty}^{\infty}\frac{dx}{\sinh x}\frac{\sinh\left((1-\Phi)x\right)}{\cosh \Phi x},\nonumber\\
\langle\sigma_i^x\sigma_{i+1}^x\rangle &=&\frac{\cos\pi\Phi}{\pi^2}\int_{-\infty}^{\infty}\frac{dx}{\sinh x}\frac{x\cosh x}{\cosh^2(\Phi x)}-\frac{1}{\pi\sin(\pi \Phi)}\int_{-\infty}^{\infty}\frac{dx}{\sinh x}\frac{\sinh\left((1-\Phi)x\right)}{\cosh \Phi x},\nonumber\\
\end{eqnarray}
where $\phi=\frac{1}{\pi}\cos^{-1}\Delta$. For $\Delta>1$, the correlation functions are given by~\cite{tks}
\begin{eqnarray}
\langle\sigma_i^z\sigma_{i+1}^z\rangle &=&1+2\int_{-\infty+i/2}^{\infty+i/2}\frac{dx}{\sinh(\pi x)}\left(\cot(\nu x)\coth(\nu)-\frac{x}{\sin^2(\nu x)}\right),\nonumber\\
\langle\sigma_i^x\sigma_{i+1}^x\rangle &=&\int_{-\infty+i/2}^{\infty+i/2}\frac{dx}{\sinh(\pi x)}\left(\frac{x}{\sin^2 (\nu x)}\cosh\nu-\frac{\cot(\nu x)}{\sinh\nu}\right),
\end{eqnarray}
where $\nu =\cosh^{-1}\Delta$. And for $\Delta\leq-1$, $\langle\sigma_i^z\sigma_{i+1}^z\rangle=1$ and $\langle\sigma_i^x\sigma_{i+1}^x\rangle=0$~\cite{ljto}. 
\begin{figure}[!ht]\centering
\includegraphics[width=6.9cm]{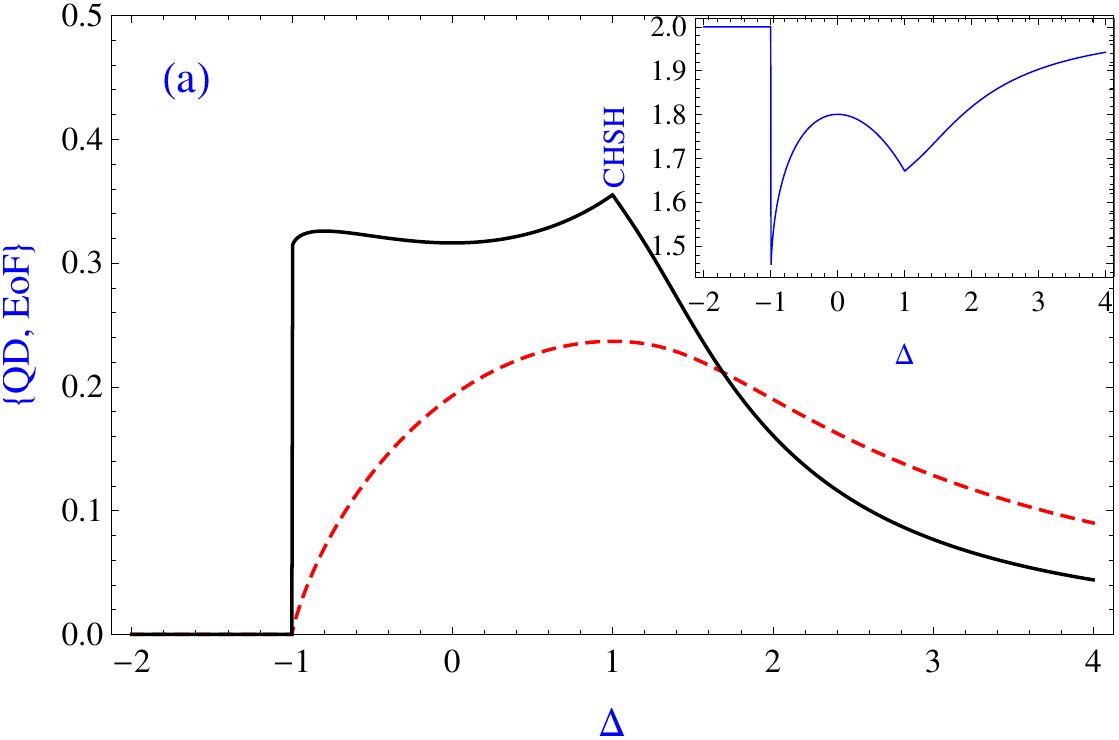}
\includegraphics[width=6.9cm]{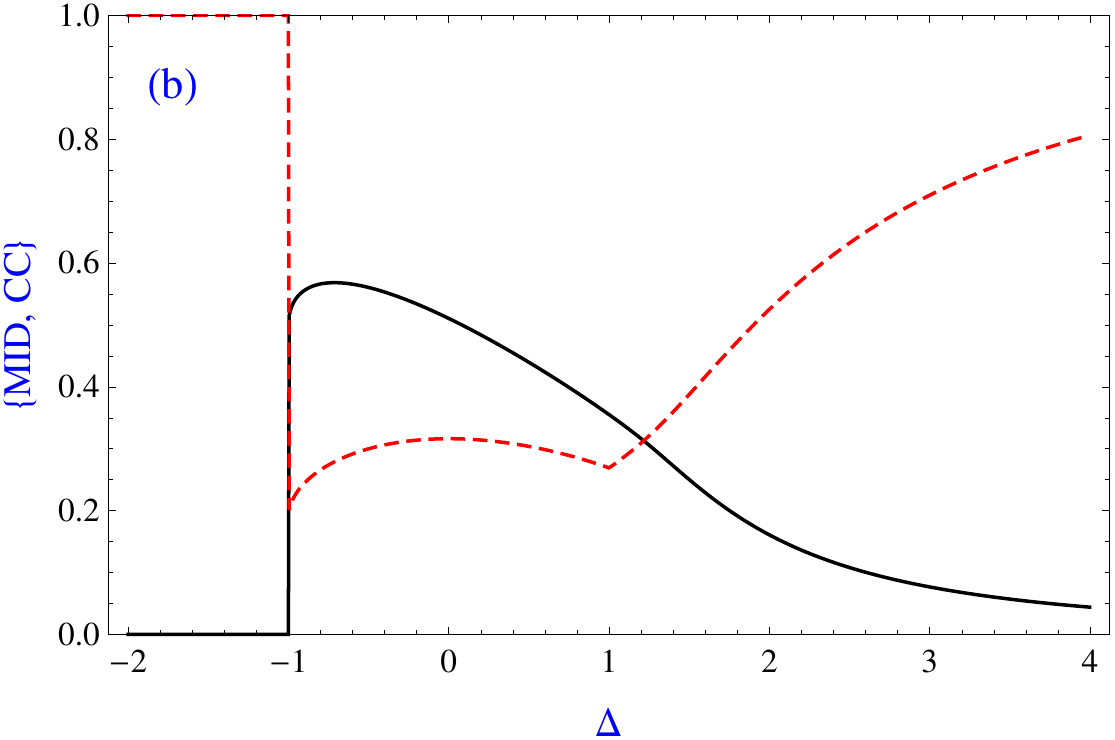}
\caption{\label{figxxz}(Color online) (a) shows EoF~(red-dashed), QD~(black-solid) and CHSH~(inset), while (b) shows CC~(red-dashed) and MID~(black-solid) for nearest-neighbor spins as a function of $\Delta$ in the XXZ chain.}
\end{figure}

In Fig.~\ref{figxxz}, we display the correlation measures as a function of anisotropy parameter $\Delta$ in the XXZ chain at $T=0$ and $N\rightarrow\infty$. At the first order critical point, $\Delta_c=-1$, all correlation measures, except EoF, are discontinuous showing the presence of the QPT, while this jump behavior is not observed in EoF. At the infinite order QPT, $\Delta_c=1$, QD, CC and CHSH function show  abrupt cusps. The cusp behavior in the considered correlation measures is due to the level-crossing in the model which involves transition from the gapless to antiferromagnetic phases. On the other hand, EoF is less sensitive to this level crossing; it just reaches its maximum at $\Delta_c=1$ which is also the case for QD~\cite{wrr,dillenschneider}. These cusplike behavior at an infinite order QPT point implies that the first derivatives of QD, CC and CHSH are discontinuous at $\Delta_c=1$, while their second derivatives are divergent. Surprisingly, MID~(black line in Fig.~\ref{figxxz}(b)) does not exhibit any abrupt cusp or an extremum point at $\Delta_c=1$. It is continuous at the phase transition point and shows no sign of the infinite order phase transition. By using the correlation functions in Refs.~\cite{ksts,tks}, we have also calculated MID for next-nearest neighbors, and have observed that MID cannot signal the infinite order phase transition for this case as well. Note that the infinite order phase transition cannot be revealed by the ground state energy of the system. Nonetheless, some of the quantum information tools (QD, CC and CHSH) can indicate this critical point with a cusplike behavior. In fact, around this critical point, there is a competition between the correlation functions $\left\langle\sigma_i^z\sigma_{i+1}^z\right\rangle$ and $\left\langle\sigma_i^x\sigma_{i+1}^x\right\rangle$ as a function of $\Delta$ and they encounter a crossover at the critical point, $\Delta_c=1$~\cite{sarandy}. Similarly, the constituents in the definitions of these correlation measures~(we mean $Q_1$ and $Q_2$ in Eq.~(\ref{qdcc}) for QD, and similar for the others) exhibit the same behavior at and around this critical point. Therefore, these cusplike behaviors can be understood due to the extremization procedure in the definitions of QD, CC and CHSH. Since MID has no extremization in its definition, it has no indication of this phase transition point. Although EoF is governed by a single non-zero variable~(since $u_{14}=0$ for the XXZ model), it reaches its maximum at this critical point. However, this is not the generic behavior of entanglement at the infinite order phase transition which may depend on the specific symmetries of the model~\cite{afov}. It is remarkable to note that between $0<\Delta<1$, QD and EoF increase with $\Delta$, while MID is a decreasing function of anisotropy in this region. On the other hand, they are all decreasing functions of $\Delta$ for $\Delta>1$ and reach zero for $\Delta\rightarrow\infty$. We should note that while MID and CC behave similarly for Ising and XY models as functions of tuning parameters, as discussed previously, they exhibit a completely different dependence on $\Delta$ for the XXZ model. See also Ref.~\cite{ljto}, where the authors showed that the CHSH-Bell function, although not violated, is found capable to detect the presence of the QPT at $\Delta_c=1$ for next- and next-next-nearest neighbor spins where entanglement is absent for such distances in the XXZ model around the critical point.
\subsection{LMG model}\label{sec4}
Finally, we analyze the ground state correlations of the model introduced by Lipkin, Meshkov and Glick~(LMG)~\cite{lmg}. The LMG model, originally, describes a two-level Fermi system with one-level below the Fermi level, while the other level is just above. The level below the Fermi level is assumed to be filled with $\Gamma$ nucleons~(here $\Gamma$ can also be thought of as a degeneracy). In its spin 1/2 representation, the LMG model can be regarded as a chain of $N$ spin particles with infinite-ranged interactions, where each spin is subject to an external transverse magnetic field $\lambda$. Indeed, the Hamiltonian of the LMG model can be written as~\cite{wsls,sarandy,prps}:
\begin{eqnarray}\label{lmghamiltonian}
H_{LMG}=-\frac{1}{N}\left(S_x^2-S_y^2\right)+\lambda S_z,
\end{eqnarray}
where $S_z=\sum_{m=1}^N\frac{1}{2}\left(c_{+m}^{\dagger}c_{+m}-c_{-m}^{\dagger}c_{-m}\right)$ and $S_x+iS_y=\sum_{m=1}^Nc_{+m}^{\dagger}c_{-m}$, with $c_{+m}^{\dagger}$ and $c_{-m}^{\dagger}$ being the particle creation operators in the upper $(+m)$ and lower $(-m)$ levels, respectively. The model has a second-order quantum phase transition at $\lambda_c=1$ between a symmetric phase~($\lambda>1$) and a broken phase~($\lambda<1$)~\cite{sarandy,rvm}. The ground state is unique and fully polarized in the direction of the field for $\lambda>1$, while it is twofold degenerate for $\lambda<1$. 

The ground state of the LMG model for $N\rightarrow\infty$ can be obtained by using the Hartree-Fock~(HF) approach. Indeed, the HF ground state provides the exact description of the critical point. The pairwise ground state density matrix for general modes $(+m)$ and $(-n)$ in the HF ground state of the LMG model can be found as~\cite{sarandy}
\begin{eqnarray}\label{lmgden}
\rho_{m,-n}=\left (\begin{array}{cccc} \left\langle M_{+m}M_{-n}\right\rangle  & 0 & 0  & 0 \\ 0  & \left\langle M_{+m}N_{-n}\right\rangle & \left\langle c_{+m}^{\dagger}c_{-n}\right\rangle  & 0 \\ 0  & \left\langle c_{-n}^{\dagger}c_{+m}\right\rangle & \left\langle N_{+m}M_{-n}\right\rangle  & 0 \\ 0  & 0 & 0  & \left\langle N_{+m}N_{-n}\right\rangle \end{array} \right) \ ,
\end{eqnarray}
where $M_k=1-N_k$ and $N_k=c_k^{\dagger}c_k$. The matrix elements can be given as
\begin{eqnarray}\label{lmgdenmatele}
\left\langle M_{+m}M_{-n}\right\rangle&=&(1-\delta_{mn})sin^2\alpha\cos^2\alpha,\nonumber\\
\left\langle M_{+m}N_{-n}\right\rangle&=&\delta_{mn}\cos^2\alpha+(1-\delta_{mn})\cos^4\alpha,\nonumber\\
\left\langle N_{+m}M_{-n}\right\rangle&=&\delta_{mn}\sin^2\alpha+(1-\delta_{mn})\sin^4\alpha,\nonumber\\
\left\langle N_{+m}N_{-n}\right\rangle&=&(1-\delta_{mn})sin^2\alpha\cos^2\alpha,\nonumber\\
\left\langle c_{+m}^{\dagger}c_{-n}\right\rangle&=&\delta_{mn}\cos\alpha\sin\alpha,
\end{eqnarray}
where $\delta_{mn}$ is the Kronecker-Delta function and  $\alpha=\frac{1}{2}\cos^{-1}\lambda$ for $\lambda<1$ and $\alpha=0$ for $\lambda\geq 1$~(see Refs.~\cite{sarandy} and~\cite{prps}, for the detailed calculations of Eqs.~(\ref{lmgden}) and~(\ref{lmgdenmatele})). One should note that the exact ground state of the LMG model, instead of the HF ground state studied here, can be obtained and highly entangled for $\lambda>1$~\cite{vidall}.
\begin{figure}[!ht]\centering
\includegraphics[width=6.4cm]{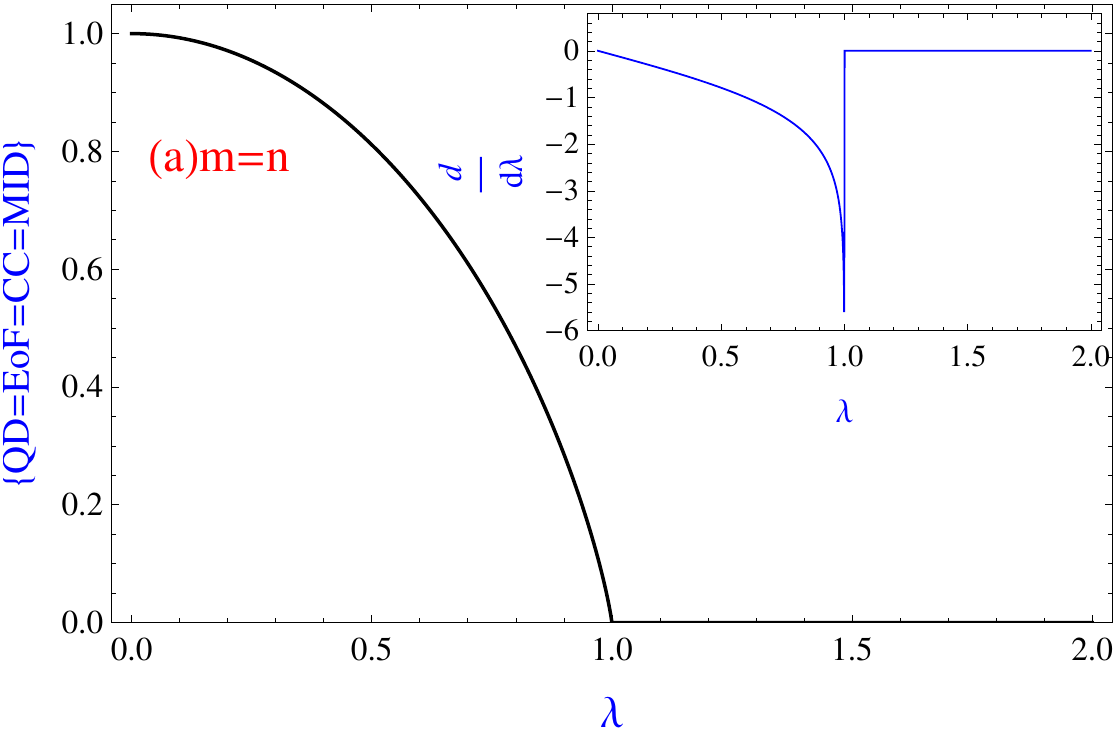}
\includegraphics[width=6.4cm]{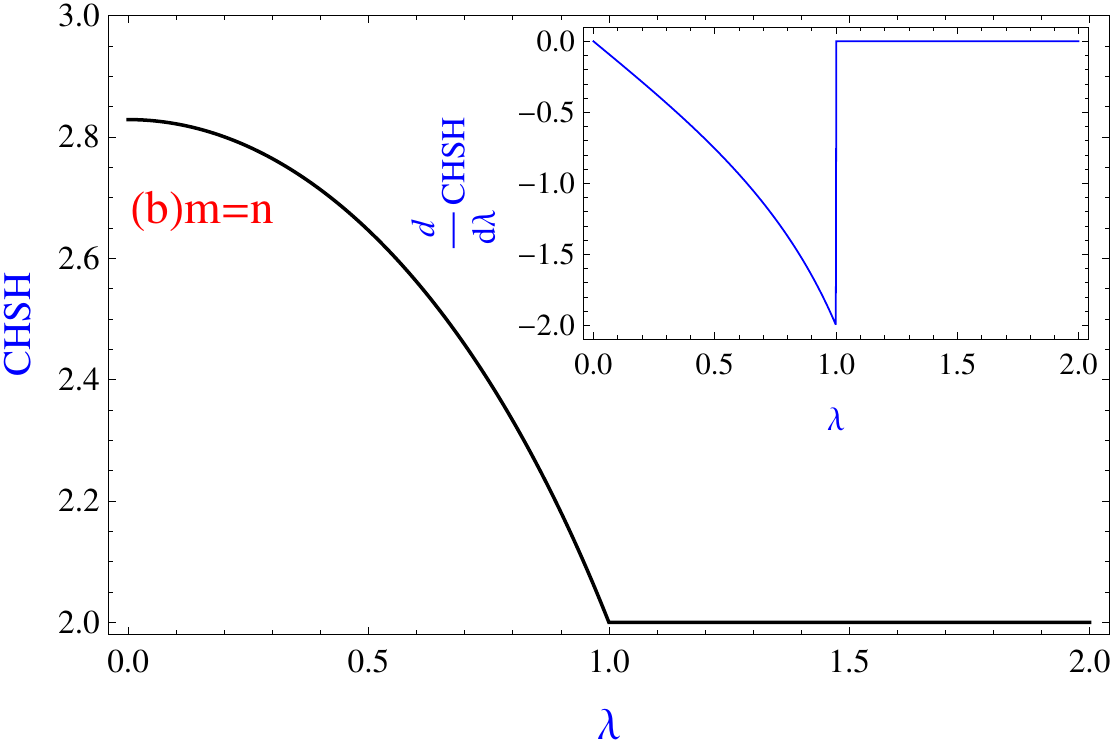}

\includegraphics[width=6.4cm]{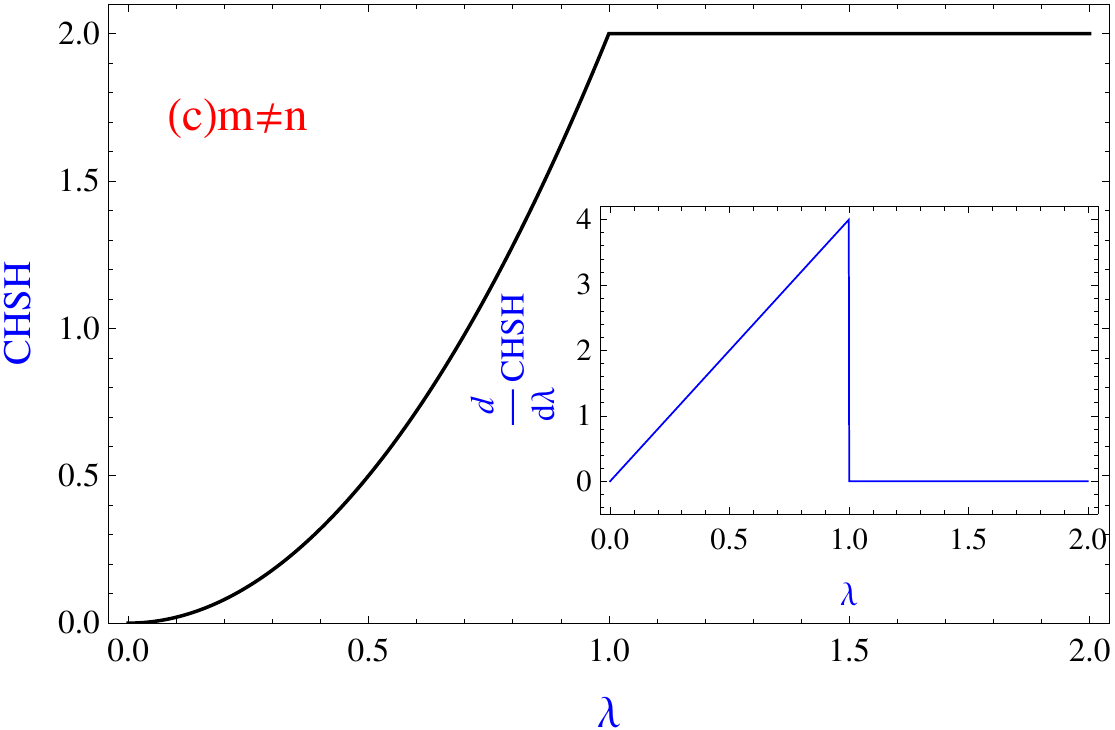}
\caption{\label{figlmg}Correlations~($QD=EoF=CC=MID$)~(a) and CHSH-Bell function~(b) between the modes~($+m$) and ($-m$) as a function of $\lambda$ in the HF ground state of the LMG model. (c) shows CHSH function for the modes $m\neq n$. Insets are the corresponding first order $\lambda$-derivative of correlation measures as a function of $\lambda$. Note that for $m\neq n$, there are no bipartite correlations, i.e., $QD=EoF=CC=MID=0$.} 
\end{figure}

In Fig.~\ref{figlmg}, we display the calculated correlations and their first derivatives with respect to $\lambda$ between the modes $(+m)$ and $(-m)$ as well as between the modes $(+m)$ and $(-n)$~($m\neq n$) in the HF ground state of the LMG model. CC, QD, MID and EoF are all equal for the density matrix of the ground state for the $(+m,-m)$ modes, because for those modes the ground state is pure and these measures are equal for the pure states. The value of correlations range from 1 for $\lambda=0$ to zero for $\lambda>1$~(see Fig.~\ref{figlmg}(a)). Also, the CHSH-Bell function is violated for $\lambda<1$~(Fig.~\ref{figlmg}(b)). The derivatives of correlation measures are discontinuous at $\lambda_c=1$, signaling the QPT~(see the insets in Figs.~\ref{figlmg}(a) and~\ref{figlmg}(b)). On the other hand, for $m\neq n$, there are no correlations~(MID, QD, CC and EoF) between the modes. In fact, the density matrix~Eq.~(\ref{lmgden}) is diagonal for $m\neq n$ and contains no pairwise classical and quantum correlations. However, the diagonal elements vary with $\lambda$ and the pairwise density matrix changes between the maximally mixed state for $\lambda=0$ and the fully polarized state for $\lambda>1$. Although this transition does not contain any bipartite correlations, remarkably, the CHSH-Bell function changes with $\lambda$ and the discontinuity~(jump) in its first derivative at $\lambda_c=1$ exhibits the QPT~(see Fig.~\ref{figlmg}(c)).

\section{Conclusions}\label{sec5}
We have investigated the quantum phase transitions of the critical systems, such as transverse field Ising, XY, XYT, XXZ and LMG models by using different quantumness measures, such as entanglement of formation, quantum discord, as well as its classical counterpart, measurement induced disturbance and CHSH-Bell function. Different aspects of quantum correlations of a given state are compared and contrasted when the tuning parameters of the system's Hamiltonians are changed, specifically at different phase regions. The ability of correlation measures to detect different order QPTs and their finite-size scaling behavior at different continuous phase transition points have been also investigated.

The most important findings of the present work can be summarized as follows. The correlation measures that we consider reflect somewhat different aspects of the quantum as well as classical properties of the ground state of the examined models and it is not possible to find a universal behavior of these measures for signaling the quantum phase transition critical point. Furthermore, for individual measures, the dependence on the tuning parameter near the critical points is not universal across the order of the transition. In particular, for the first order QPT observed in XXZ model, all the considered measures, except EoF, show a discontinuity behavior at the CP, while at the infinite-order QPT observed in the same model, MID has no indication of the QPT.  On the other hand, EoF has its maximum and QD, CC and CHSH have a cusp-like structure, showing the infinite-order phase transition. The number of second-order QPTs investigated in the present work is much higher than that of the first or infinite-order QPTs. None of the considered measures was found to behave exactly the same at CP for the investigated seven second-order phase transitions. Except for the LMG model for $(+m,-n)$ modes with $m\neq n$, the first or second derivatives of all correlation measures are found to be discontinuous or divergent at the critical point of the relevant models, signaling the CPs of QPTs. Remarkably, the CHSH function is the sole indicator of the phase transition between $(+m)$ and $(-n)$ modes for $m\neq n$ in the HF ground state of the LMG model. Moreover, the tuning parameter dependence of MID is found to be completely different to that of QD in some phase regions. Interestingly, for Ising and XY models, the classical correlations and MID as functions of tuning parameters are nearly the same.

In fact, the main motivation of this study was to search for new quantum information theory based indicators of quantum phase transitions, such as MID and the CHSH-Bell function, and compare them with the well known CP-detectors, QD, CC and EoF. It is found that the finite-size scaling behaviors of MID and CHSH-Bell function in Ising and XYT models exhibit a logarithmic scaling which is similar to that of EoF, QD and CC. On the other hand, MID cannot  signal the infinite-order phase transition present in the XXZ model, although the others can.
Above all, first and foremost, the CHSH-Bell function is found to exhibit all the phase transitions present in the considered critical models, even when the relevant ground state is pairwise quantum and classical uncorrelated. This result is, indeed, very desirable for the quantum information community, because the definition or analytic forms of quantum correlation measures for mixed states are mostly given for bipartite systems. Although the unbroken bipartite state of the entire system is expected to provide the exact description of the critical behavior as well as its scaling in finite systems, sometimes the bipartite density matrix can loose its coherence, and thus, QD, CC and EoF can become insufficient to detect the CPs of QPTs, such as for the ($+m$,$-n$) mode case of the LMG model in HF ground state, for very far neighbor case and at finite temperatures~\cite{wsls,wtrr,ylhl,plaqpt}. On the other hand, the CHSH-Bell function does not frozen until the bipartite state is frozen. Moreover, the definition and also the analytic form (for some special cases) of CHSH-Bell function for multipartite mixed states are available. We believe that the definition and the non-ergodic behavior of the non-locality measure make it more sophisticated CP-detector than the other well known CP-detectors and MID. However, its ability to detect the CPs of QPTs in different critical systems and particularly at finite temperatures needs strict investigations to make a direct connection between CHSH function as a CP-detector and experiments. Also, it is interesting to note that the bipartite states~(except the LMG model for the $(+m,-m)$ mode case) are mixed and do not violate the CHSH inequality, although they contain entanglement. This indicates that the correlations possessed by the two-point ground states are highly local. 

The quantum correlation measures are not expected to signal the classical phase transitions, since the two-point density matrix can lose its coherence when the temperature rises~\cite{mgcss,wtrr,wrr,ylhl}. Nonetheless, very recently, QD was found to be able to indicate the critical temperature, $T_c$, of the classical phase transitions in the derived model from the two-dimensional square Ising model, while entanglement could not~\cite{cptin}. It would be desirable to investigate whether or not the CHSH-Bell function can signal the CPs of classical phase transitions. It would be also very interesting to study different quantumness measures~(see the recent review article~\cite{revar}), especially the ameliorated MID to see whether or not this amplified version of MID can signal the infinite order phase transition in the XXZ model. Also, the analysis to see whether or not this is a generic feature of MID at infinite order phase transitions exhibited by different models~\cite{ccrmss,crms} needs further consideration.

{\bf Note added:} After finishing this work we became aware of two related studies~\cite{rework1,rework2} where the authors studied the performance of different quantum information 
theoretic correlation measures to indicate CPs of QPTs.

\section*{Acknowledgments}

We would like to thank anonymous Referee for constructive remarks.


\begin{thebibliography}{99}
\bibitem{orderqpt} The order of the QPTs can be determined by looking at the non-analyticities of the derivatives
of the ground state energy. If the first derivative of the ground state energy with respect
to the tuning parameter presents a finite discontinuity then the QPT is of the first order.
Similarly, if the first derivative of the ground state energy is continuous, while its second
derivative exhibits a finite discontinuity or divergence, then the QPT is second order (also
called continuous QPT). On the other hand, if any finite-order derivatives of the ground
state energy is a continuous function with respect to the relevant tuning parameter, then
the QPT is called an infinite order.
\end{thebibliography}
\end{document}